\documentclass[10pt,conference]{IEEEtran}
\IEEEoverridecommandlockouts

\usepackage{cite}
\usepackage{amsmath,amssymb,amsfonts}
\usepackage{algorithmic}
\usepackage{graphicx}
\usepackage{textcomp}
\usepackage{xcolor}
\usepackage[table]{xcolor}
\usepackage{booktabs}
\usepackage{multirow}
\usepackage{arydshln}
\usepackage{pifont}
\usepackage{tcolorbox}
\usepackage{listings}
\usepackage{float}
\usepackage{newfloat}
\usepackage{xcolor}
\usepackage{tikz}
\usepackage{subcaption} 
\usepackage{makecell}
\usepackage{lipsum}
\usepackage{cleveref}
\usepackage{mdframed}
\usepackage{url}

\definecolor{codegreen}{rgb}{0,0.6,0}
\definecolor{codegray}{rgb}{0.5,0.5,0.5}
\definecolor{codepurple}{rgb}{0.58,0,0.82}
\definecolor{backcolour}{rgb}{0.95,0.95,0.92}

\newfloat{listing}{tb}{lst}{}
\floatname{listing}{Listing}

\def\BibTeX{{\rm B\kern-.05em{\sc i\kern-.025em b}\kern-.08em
    T\kern-.1667em\lower.7ex\hbox{E}\kern-.125emX}}
\begin{document}

\title{A Multi-Language Object-Oriented Programming Benchmark for Large Language Models\\
}


\author{
\IEEEauthorblockN{
Shuai Wang\IEEEauthorrefmark{1},
Liang Ding\IEEEauthorrefmark{2},
Li Shen\IEEEauthorrefmark{3},
Yong Luo\IEEEauthorrefmark{1},
Han Hu\IEEEauthorrefmark{4},
Lefei Zhang\IEEEauthorrefmark{1},
Fu Lin\IEEEauthorrefmark{1}
}
\IEEEauthorblockA{\IEEEauthorrefmark{1}Wuhan University \quad
\IEEEauthorrefmark{2}The University of Sydney \quad
\IEEEauthorrefmark{3}Sun Yat-sen University \quad
\IEEEauthorrefmark{4}Beijing Institute of Technology}
\IEEEauthorblockA{Emails: wangshuai123@whu.edu.cn, liangding.liam@gmail.com}
}

\maketitle

\begin{abstract}
Establishing fair and robust benchmarks is essential for evaluating intelligent code generation by large language models (LLMs). Our survey of 35 existing benchmarks uncovers three major imbalances: 85.7\% focus on a single programming language; 94.3\% target only function-level or statement-level tasks; and over 80\% include fewer than ten test cases on average. 

To address these gaps, we propose MultiOOP, a multi-language object-oriented programming benchmark covering six popular languages (Python, PHP, C++, C\#, Java, JavaScript) with 267 tasks per language. We design a translator that extends an existing single-language OOP benchmark and the \textit{pass@$o$} metric to a multilingual setting. Moreover, we propose an automated framework for augmenting test cases to ensure the reliability of the evaluation results.
We evaluate 14 mainstream LLMs under zero-shot prompting and report three key findings:
1) \textit{Substantial performance degradation}: \textit{pass@$1$} scores on MultiOOP drop by up to 65.6 percentage points compared to function-level tasks (e.g., HumanEval).
2) \textit{Cross-language variability}: GPT-4o mini achieves \textit{pass@$1$} of 48.06\% in Python but only 0.12\%–15.26\% in other languages, indicating limited multilingual generalization.
3) \textit{Conceptual gaps}: \textit{pass@$o$} scores are consistently 1.1–19.2 points lower than \textit{pass@$k$}, demonstrating that LLMs often generate executable code without fully capturing core OOP concepts.
Our benchmark, metric extensions, and evaluation scripts will be publicly released to foster a more balanced and comprehensive assessment of LLMs in object-oriented code generation\footnote{Our code and data will be released at \url{https://github.com/alphadl/OOP-eval} and \url{https://huggingface.co/datasets/codeai-dteam/MultiOOP} respectively.}.
\end{abstract}

\begin{IEEEkeywords}
Code Generation, Large Language Models, Object-Oriented Programming, Multilingual Benchmarks.
\end{IEEEkeywords}

\section{Introduction}
Traditional software development and system design~\cite{herbsleb2003empirical,mockus2002two} primarily involve manual coding, which is labor-intensive and error-prone. Intelligent code generation tackles these issues by leveraging algorithms to transform user requirements into executable code, thereby reducing manual effort and improving both efficiency and quality~\cite{qian2024chatdev,yang2024multi}.


\begin{table}[!t]
  \centering
  \caption{\textbf{Overview of existing code generation evaluation benchmarks}. (``NL'' denotes natural language describing the problem or requirements; ``PL'' represents the generated programming language; ``Multi'' means containing multiple NLs or PLs; ``$\star$'' indicates the number of samples for each language; Gray indicates benchmarks for multiple programming languages.)}
  \resizebox{1.0\linewidth}{!}{
    \begin{tabular}{l|clllll}
    \toprule
    Benchmark & Time & Number & NL    & PL    & AvgTest & Task Granularity \\
    \midrule
    Concode~\cite{iyer2018mapping} & 2018  & 2000  & en    & Java  & - & Function \\
    CoNaLA~\cite{yin2018learning} & 2018  & 500   & en    & Python & - & Statement \\
    HumanEval~\cite{chen2021evaluating} & 2021  & 164   & en    & Python & 7.7   & Function \\
    MBPP~\cite{austin2021program}  & 2021  & 974   & en    & Python & 3.0     & Function \\
    APPS~\cite{hendrycks2021measuring}  & 2021  & 5000  & en    & Python & 13.2  & Function \\
    CodeContests~\cite{li2022competition} & 2022  & 165   & en    & Python & 203.7 & Function \\
    DSP~\cite{chandel2022training}   & 2022  & 1119  & en    & Python & 2.1   & Function \\
    \rowcolor[rgb]{ .906,  .902,  .902} MBXP-HumanEval~\cite{athiwaratkun2023multi} & 2022  & $164^{\star}$ & en    & Multi & 7.8   & Function \\
    \rowcolor[rgb]{ .906,  .902,  .902} MultiPL-HumanEval~\cite{cassano2023multipl} & 2022  & $164^{\star}$ & en    & Multi & 7.8   & Function \\
    \rowcolor[rgb]{ .906,  .902,  .902} MultiPL-MBPP~\cite{cassano2023multipl} & 2022  & 974   & en    & Multi & 3.0     & Function \\
    PandasEval~\cite{zan2022cert} & 2022  & 101   & en    & Python & 6.5   & Function \\
    NumpyEval~\cite{zan2022cert} & 2022  & 101   & en    & Python & 3.5   & Function \\
    TorchDataEval~\cite{zan2022language} & 2022  & 50    & en    & Python & 1.1   & Function \\
    \rowcolor[rgb]{ .906,  .902,  .902} HumanEval-X~\cite{zheng2023codegeex} & 2023  & $164^{\star}$ & en    & Multi & 7.8   & Function \\
    HumanEval$^{+}$~\cite{liu2024your} & 2023  & $164$ & en    & Python & 764.1   & Function \\
    MTPB~\cite{liu2024multi}  & 2022  & 115   & en    & Python & - & Function \\
    BIG-Bench~\cite{srivastava2022beyond} & 2022  & 32    & en    & Python & 4.7   & Function \\
    ODEX~\cite{wang2023execution}  & 2022  & 945   & Multi & Python & 1.8   & Function \\
    CodeApex~\cite{fu2023codeapex} & 2023  & 476 & en\&zh & C++   & - & Function \\
    DS-1000~\cite{lai2023ds} & 2022  & 1000  & en    & Python & 1.6   & Statement \\
    \rowcolor[rgb]{ .906,  .902,  .902} CoderEval & 2023  & $230^\star$   & en    & Python\&Java & - & Function \\
    ClassEval~\cite{yu2024codereval} & 2023  & 100   & en    & Python & 33.1  & Class \\
    TACO~\cite{li2023taco}  & 2023  & 1000  & en    & Python & 0     & Function \\
    OOP~\cite{wang-etal-2024-oop}   & 2024  & 431   & en    & Python & 2.5   & Object-oriented \\
    JavaBench~\cite{cao2024javabench} & 2024  & 106   & en    & Java  & 99    & Object-oriented \\
    EvoCodeBench~\cite{lievocodebench} & 2024  & 275   & en    & Python & 3.5   & Repository \\
    PECC~\cite{haller2024pecc} & 2024  & 2352   & en    & Python & -   & Function \\
    Exec-CSN~\cite{xie2024codebenchgen} & 2024  & 1931  & en    & Python & - & Repository \\
    StudentEval~\cite{babe2023studenteval} & 2024  & 1749  & en    & Python & 3.2     & Function \\
    DSEval~\cite{zhang2024benchmarking} & 2024  & 825  & en    & Python & -     & Function \\
    EvoCodeBench~\cite{li2024evocodebench} & 2024  & 275  & en    & Python & -     & Repository \\
    DA-Code~\cite{huang2024code} & 2024  & 500  & en    & Python & -     & Statement \\
    BigCodeBench~\cite{zhuo2024bigcodebench} & 2024  & 1140  & en    & Python & 5.6     & Statement \\
    LiveCodeBench~\cite{jain2024livecodebench} & 2024  & 511  & en    & Python & 17.1     & Statement \\
    LeetCodeDataset~\cite{xia2025leetcodedataset} & 2025  & 2870  & en    & Python & 100+     & Function \\
    \toprule
     \textbf{MultiOOP (our)} & \textbf{2024}  & \textbf{$267^\star$}   & \textbf{en}    & \textbf{Multi} & \textbf{20.0}     & \textbf{Object-oriented} \\
    \bottomrule
    \end{tabular}%
    }
  \label{tab:exist_benchmark}%
\end{table}%

In recent years, with the rapid advancement of natural language processing (NLP) technology~\cite{hirschberg2015advances,otter2020survey,zhong2022toward,zan2022vega,treviso2023efficient,zhong2023can,peng2023towards}, particularly the emergence of large language models (LLMs) like GPT~\cite{achiam2023gpt} and Llama~\cite{touvron2023llama,dubey2024llama}, the LLMs have been trained on extensive programming corpora, incorporating a vast number of code samples. This allows them to understand the syntax and common programming patterns across different programming languages. 
In the intelligent code generation process, the input to LLMs typically consists of natural language descriptions or partial code snippets.
By understanding the input, LLMs can generate code that is both syntactically correct and logically sound.
For example, when a user inputs ``write a bubble sort algorithm using the Python programming language'', the LLM understands the user's requirements and generates the corresponding sorting code.

To objectively and fairly assess the code generation capabilities of emerging LLMs, many studies~\cite{chen2021evaluating,austin2021program,li2022competition,hendrycks2021measuring,athiwaratkun2023multi,wang-etal-2024-oop} have focused on evaluating LLMs using automatically or manually constructed code generation benchmarks, e.g., HumanEval~\cite{chen2021evaluating} and MBPP~\cite{austin2021program}. 
Although existing benchmarks are valuable for evaluating and comparing the performance of LLMs, most focus primarily on a single language and are limited to function-level or statement-level tasks. Additionally, they typically include a relatively low number of test cases, as illustrated in Table~\ref{tab:exist_benchmark}.


\textbf{[Limitations of existing benchmarks]} From Table~\ref{tab:exist_benchmark} in the survey, we can observe three notable imbalances: 1) Imbalance in programming languages. More than 85.71\% (30/35) of benchmarks focus on a single-language, with four supporting multiple languages and one covering two languages. 2) Imbalance in task granularity. 
Over 94.29\% (33/35) of the benchmarks concentrate on function-level, statement-level, or library-level programming, with only two focusing on OOP. As both benchmarks are limited to a single programming language, they fail to adequately capture LLMs’ multilingual code generation capabilities.
3) Imbalance in test cases. The average number of test cases exceeding 80.00\% (28/35) of the benchmark is fewer than 10, which does not adequately ensure the reliability of evaluation results.


\textbf{[Motivation]} 
It is well known that OOP is more efficient, flexible, and reliable in software development due to its characteristics, e.g., modularity, encapsulation, inheritance, polymorphism, scalability, and maintainability~\cite{briot1998concurrency,moon1986object,madsen1989virtual,wilde1992maintenance}. However, practical software development is not limited to a single language. Therefore, establishing a multi-language object-oriented programming (\textbf{MultiOOP}) evaluation framework with multiple test cases to evaluate the generalization capabilities of LLMs across different programming language scenarios has become an urgent necessity.

\textbf{[MultiOOP benchmark and metric]} To fill this gap, we develop a programming language translator that converts a single-language OOP benchmark~\cite{wang-etal-2024-oop} into a multilingual one. The translated benchmark (namely MultiOOP) includes OOP tasks in multiple languages (i.e., Python, PHP, C++, C\#, Java, and JavaScript), with 267 samples for each language. These OOP tasks cover fundamental concepts such as classes, inheritance, and encapsulation. 
Moreover, we extend the traditional \textit{pass@$o$} metric~\cite{wang-etal-2024-oop} to the multilingual setting to enhance its generalizability in OOP tasks. Additionally, we systematically design a framework for automatically generating test cases for MultiOOP tasks, ensuring the reliability of evaluation results.

\textbf{[Evaluation results]} In our evaluation of 14 mainstream LLMs, we identified several notable findings:
1) Current LLMs exhibit significantly lower performance on MultiOOP generation compared to function-level or statement-level programming tasks (e.g., HumanEval~\cite{chen2021evaluating}, MBPP~\cite{austin2021program}).
2) There is considerable variation in the performance of existing LLMs on OOP tasks across different programming languages.

\textbf{[Contributions]} Our main contributions are as follows:
\begin{itemize}
\item We have carefully designed a programming language translator that can convert a single-language OOP benchmark into a multi-language benchmark, and is suitable for unit tests, test cases, etc.

\item We use a custom-designed programming language translator to adapt the OOP benchmark and the \textit{pass@$o$} metric for multiple languages. The extended OOP benchmark (namely MultiOOP) covers core OOP concepts, such as classes, inheritance, polymorphism, and encapsulation, with 267 samples per language. Moreover, The extended evaluation metric \textit{pass@$o$} exhibits strong generalizability across multiple languages in OOP tasks.

\item We develop a framework for automatically generating test cases for MultiOOP tasks. This framework produces test cases to verify the code generated by LLMs, thereby enhancing the accuracy and reliability of standard evaluation methods.

\item Through extensive evaluation of 14 leading LLMs on MultiOOP tasks, we show that: 1) there is considerable room for improvement in the performance of LLMs within the MultiOOP context; 2) our benchmark provides a robust and fair metric that enables the coding community to accurately assess the performance of LLMs in OOP tasks across various programming languages.
\end{itemize}

\section{Related work}
\subsection{Code generation}
Currently, LLMs demonstrate significant progress in code tasks, particularly in code generation tasks~\cite{dong2024self,wang2023review,wang2024mathbb}.
Trained on vast amounts of code and natural language data, these LLMs can automatically generate high-quality code snippets. For instance, Codex~\cite{chen2021evaluating} and ChatGPT~\cite{achiam2023gpt}, developed by OpenAI, as well as AlphaCode~\cite{li2022competition} and Gemini~\cite{team2023gemini,team2024gemini} from DeepMind, have demonstrated impressive capabilities in both code generation and comprehension. ChatGPT, originally an extension of GPT-3~\cite{brown2020language}, can interpret natural language requirements and generate corresponding code.

Based on common applications of LLMs in code generation, existing tasks can generally be categorized into two types. The first involves providing the LLM with a natural language description of the requirements, expecting it to generate the corresponding code from scratch. The second involves supplying the LLM with both a code skeleton and the requirement description, asking it to complete the implementation within the given structure, as shown in Figure~\ref{fig:human_mbpp}.


\subsection{Code evaluation benchmark}
In recent years, code evaluation benchmarks have attracted significant attention and have become essential tools for assessing the performance of LLMs in various code-related tasks, including code generation, comprehension, etc. 

Early code generation evaluation benchmarks (e.g., Concode~\cite{iyer2018mapping} and CoNaLA~\cite{yin2018learning}) primarily utilized traditional natural language processing metrics like BLEU~\cite{papineni2002bleu}, ROUGE~\cite{lin2005recall}, and METEOR~\cite{banerjee2005meteor}, which focus on the textual similarity between generated code and reference code. However, given the semantic nature of code, the metrics have notable limitations in evaluating code quality. For example, two syntactically distinct yet functionally equivalent code snippets may receive low BLEU scores. To tackle this issue, several code generation evaluation benchmarks, e.g., HumanEval~\cite{chen2021evaluating} and MBPP~\cite{austin2021program}, have been introduced, emphasizing tasks like code generation and understanding. 
Notably, HumanEval introduces a unit test-based dynamic evaluation method (namely \textit{pass@$k$}), which uses predefined test cases to assess the semantic correctness of generated code. While HumanEval addresses several challenges in code evaluation, it still has limitations, such as the difficulty of developing high-quality test cases, the difficulty in reducing bias toward specific programming languages, and the limited scope of scenarios covered by code generation tasks.
Subsequent code generation evaluation benchmarks have focused on addressing these three key challenges. For instance, MBXP-HumanEval~\cite{athiwaratkun2023multi} and MultiPL-HumanEval~\cite{cassano2023multipl} extend the HumanEval and MBPP benchmarks to support multiple languages. HumanEval$^{+}$~\cite{liu2024your} enhances the original HumanEval benchmark by adding more test cases.
Benchmarks, e.g., OOP~\cite{wang-etal-2024-oop} and JavaBench~\cite{cao2024javabench}, specifically target object-oriented programming scenarios, while others, such as PandasEval and NumpyEval~\cite{zan2022cert}, focus on Python library-specific tasks.
We have summarized these well-studied code generation benchmarks in Table~\ref{tab:exist_benchmark}.

Although existing benchmarks have advanced intelligent code generation, most focus on a single-language and overlook multilingual scenarios. Benchmarks like MBXP-HumanEval extend HumanEval and MBPP to multiple languages but mainly target function-level programming and include only a few test cases. Thus, a dedicated benchmark for MultiOOP tasks is urgently required.

\begin{figure}[!t]
\begin{subfigure}{\columnwidth}
\begin{lstlisting}[escapeinside={(*}{*)}, basicstyle=\scriptsize\ttfamily]
(*\colorbox{red!20}{\# \textbf{Code Skeleton}}*)
def truncate_number(number: float) -> float:
(*\colorbox{red!20}{\# \textbf{Requirement Description}}*)
    """Given a positive floating point number, it can be decomposed into and integer part (largest integer smaller than given number) and decimals (leftover part always smaller than 1). Return the decimal part 
     of number.
(*\colorbox{red!20}{\# \textbf{Input-output example}}*)
    >>> truncate_number(3.5)
    0.5
    """    
\end{lstlisting}
\caption{HumanEval benchmark}
\label{humaneval}
\end{subfigure}

\begin{subfigure}{\columnwidth}
\begin{lstlisting}[escapeinside={(*}{*)},basicstyle=\scriptsize\ttfamily]
You are an expert Python programmer, and here is
your task:
...
(*\colorbox{red!20}{\# \textbf{Requirement Description}}*)
Write a function to check if the given tuple list has all 
k elements.
(*\colorbox{red!20}{\# \textbf{Input-output example}}*)
Your code should pass these tests:
["assert check_k_elements([(4, 4), (4, 4, 4), 
(4, 4), (4, 4, 4, 4), (4, )], 4) == True", ...]
[BEGIN]
\end{lstlisting}
\caption{MBPP benchmark}
\label{mbpp}
\end{subfigure}
\caption{An example from the HumanEval benchmark~\cite{cassano2023multipl} and the MBPP benchmark~\cite{austin2021program}. The HumanEval benchmark task includes both code skeletons and requirement descriptions, while the MBPP benchmark only includes requirement descriptions.}
\label{fig:human_mbpp}
\end{figure}

\section{The MultiOOP evaluation framework}
\subsection{Overview}
Existing code generation evaluation benchmarks focus on single-languages and function-level or statement-level programming while largely overlooking multi-language and object-oriented programming (OOP). 
As highlighted in Table~\ref{tab:exist_benchmark}, we refer to the two most widely used benchmarks among researchers—HumanEval and MBPP, illustrated in Figure~\ref{fig:human_mbpp}. Notably, both benchmarks are constrained to function-level and statement-level evaluations, and are exclusively designed for the Python language. Moreover, the average number of test cases per problem is relatively low. Therefore, there is an urgent need to develop a benchmark that focuses on MultiOOP tasks and includes multiple test cases.


\begin{figure}[!t]
\begin{subfigure}{\columnwidth}
\begin{lstlisting}[escapeinside={(*}{*)},basicstyle=\scriptsize\ttfamily]
(*\colorbox{red!20}{\# \textbf{Code Skeleton}}*)
...
public class CellStack {
       ...
       private int count = 0; 
(*\colorbox{red!20}{\# \textbf{Requirement Description}}*)
       /** 
         *Pushes a cell into the stack. 
         *@param cell Cell to push into the stack. 
         */ 
(*\colorbox{red!20}{\# \textbf{Code Skeleton}}*)
        void push(final FillableCell cell) { 
        // Implement the requirements here.
        } 
       ...
}   
\end{lstlisting}
\caption{JavaBench benchmark (Java language)}
\label{javabench}
\end{subfigure}
\begin{subfigure}{\columnwidth}
\begin{lstlisting}[escapeinside={(*}{*)},basicstyle=\scriptsize\ttfamily]
(*\colorbox{red!20}{\# \textbf{Requirement Description}}*)
Question: Given an integer (*\textbf{n}*), please find and return the 
(*\textbf{n}*)-th ugly number. Please design a (*\textbf{ULYNB}*) class in Python
language based on the above question. The class should 
have an instance attribute (*\textbf{n}*), a private function 
(*\textbf{private\_ugly\_number}*), and a public function 
(*\textbf{public\_ugly\_number}*). In the private function 
(*\textbf{private\_ugly\_number}*), find the (*\textbf{n}*)-th ugly number based on 
the instance attribute (*\textbf{n}*). Finally, in the public function 
(*\textbf{public\_ugly\_number}*), call the private function 
(*\textbf{private\_ugly\_number}*) and return the result.
\end{lstlisting}
\caption{OOP benchmark (Python language)}
\label{oop}
\end{subfigure}
\caption{An example of a single-language object-oriented programming benchmark. The JavaBench benchmark~\cite{cao2024javabench} task includes both code skeletons and requirement descriptions, while the OOP benchmark~\cite{wang-etal-2024-oop} only includes requirement descriptions.}
\label{fig:JavaBench_OOP}
\end{figure}

\begin{figure*}[!t]
    \centering
    \includegraphics[width=0.91\textwidth]{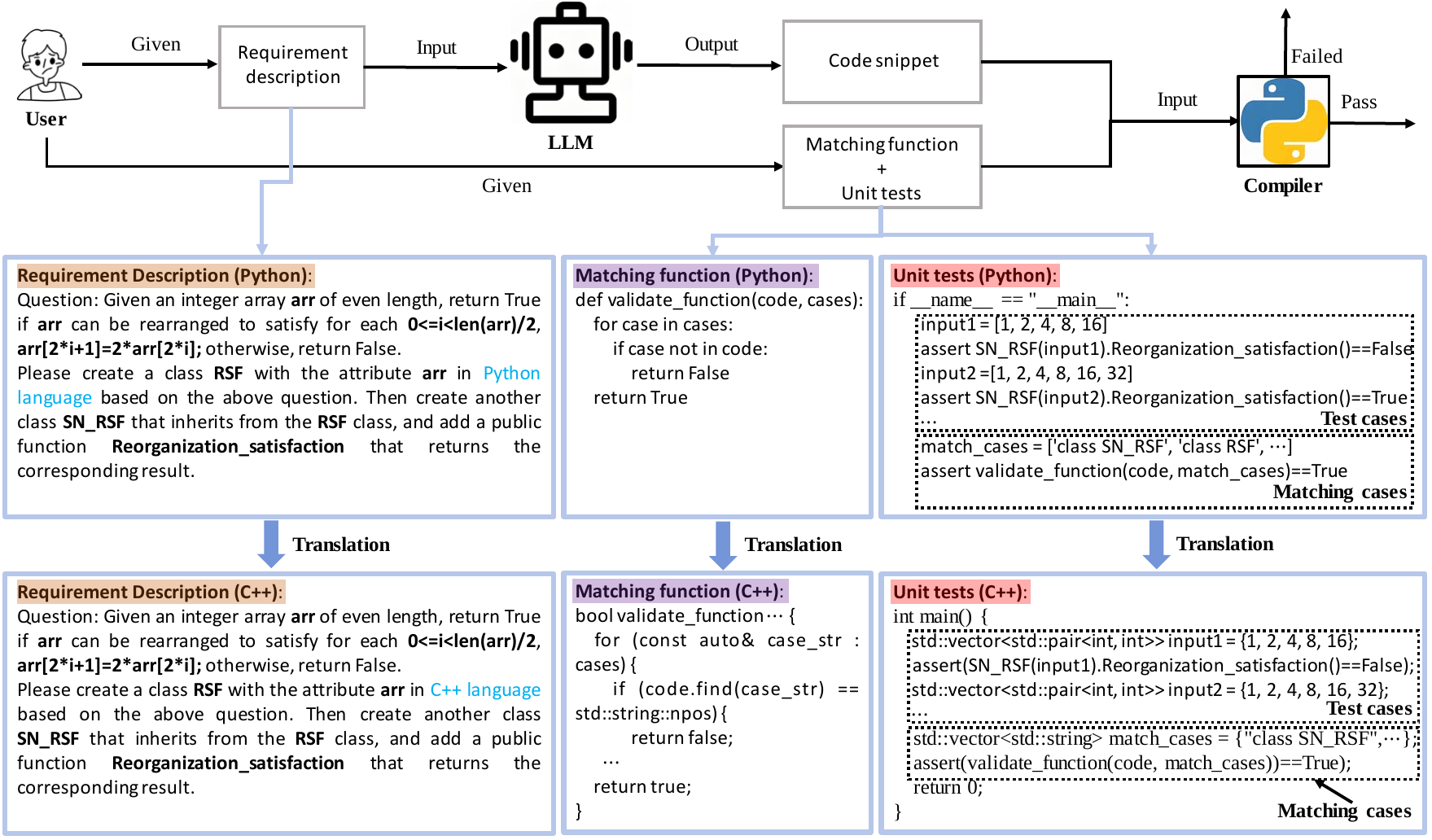}
    \caption{The process of constructing a MultiOOP evaluation benchmark (Here, we take C++ as reference. The same is true when translating Python to other programming languages i.e., Java, C\#, PHP, and JavaScript. The construction of the MultiOOP benchmark is mainly divided into three stages: translation of requirement description, unit tests, and matching function).}
    \label{fig:ovall_frame}
\end{figure*}

\subsection{Benchmark selection}
It is widely acknowledged that the pre-training corpora of existing code LLMs are primarily sourced from online code repositories such as GitHub and Stack Overflow. Consequently, evaluating these LLMs using test data from the same sources poses a risk of training–test overlap, potentially leading to data contamination. Such contamination can significantly compromise the objectivity and fairness of the evaluation process, thereby undermining a reliable assessment of LLMs' generalization capabilities in OOP scenarios.

Therefore, we plan to translate the existing single-language object-oriented programming benchmarks into multiple languages to prevent data contamination. According to our survey in Table~\ref{tab:exist_benchmark}, there are currently two single-language object-oriented programming benchmarks, namely the OOP benchmark~\cite{wang-etal-2024-oop} and the JavaBench benchmark~\cite{cao2024javabench}, as illustrated in Figure~\ref{fig:JavaBench_OOP}. Among them, the OOP benchmark is a good choice for several reasons. 

First, the OOP benchmark comprises 431 independent samples, each featuring input in the form of natural language requirement descriptions. In contrast, JavaBench places greater emphasis on implementing specific functionalities within a predefined object-oriented structure, as illustrated in Figure~\ref{fig:JavaBench_OOP}.
In comparison, the OOP benchmark is more suited for evaluating the ability of LLMs to automatically generate complete OOP algorithms. 
Secondly, in real-world software development, beginners often express functional requirements in natural language, which makes the scenario modeled by the OOP benchmark more realistic. Consequently, the results obtained from the OOP benchmark offer more valuable insights for users with limited programming experience.
Finally, due to the inherent differences between programming languages (see section~\ref{sec:translation_OOP}), translating code skeletons in the JavaBench benchmark may introduce additional errors. In contrast, the OOP benchmark, which uses natural language input descriptions, requires little to no translation or structural transformation of code skeletons.

To this end, we carefully designed a translator that can convert OOP benchmark into multiple languages, thereby constructing a MultiOOP benchmark, as shown in Figure~\ref{fig:ovall_frame}. 
Based on user code generation and evaluation with LLMs, our translation workflow comprises three main stages: translating requirement description, unit tests, and matching function (see section~\ref{sec:translation_OOP}). Additionally, the translation of unit tests and matching function extends the evaluation \textit{pass@$o$} metric.

\begin{table}[!t]
  \centering
  \caption{The ranking of six programming languages (i.e., C++, Java, C\#, PHP, Python, and JavaScript) across four major platforms (i.e., TIOBE, GitHut, PYPL, and RedMonk.}
  \resizebox{0.98\linewidth}{!}{
    \begin{tabular}{cccccc}
    \toprule
    PL    & \multicolumn{1}{c}{GitHut 2.0} & \multicolumn{1}{c}{TIOBE} & \multicolumn{1}{c}{PYPL} & \multicolumn{1}{c}{RedMonk} & \multicolumn{1}{c}{Category} \\
    \midrule
    C++   & 9.49\% & 2     & 6.90\% & 8     & Medium \\
    Java  & 11.75\% & 3     & 15.52\% & 3     & High \\
    C\#   & 3.45\% & 5     & 6.48\% & 5     & Low \\
    PHP   & 5.69\% & 12    & 4.09\% & 4     & Low \\
    Python & 16.99\% & 1     & 29.39\% & 2     & High \\
    JavaScript & 9.90\% & 6     & 8.16\% & 1     & Medium \\
    \bottomrule
    \end{tabular}%
    }
  \label{tab:programm_language}%
\end{table}%

\begin{table*}[!t]
  \centering
  \caption{The translation content of unit tests.}
  \resizebox{1.0\linewidth}{!}{
    \begin{tabular}{ccccccc}
    \toprule
    \multicolumn{1}{c}{Translate the converted content} & Python & C++   & Java  & C\#   & PHP   & JavaScript \\
    \midrule
    \multicolumn{7}{c}{Non object-oriented programming} \\
    \midrule
    Declaration of variable types & \multicolumn{1}{c}{\ding{55}} & \multicolumn{1}{c}{\checkmark} & \multicolumn{1}{c}{\checkmark} & \multicolumn{1}{c}{\checkmark} & \multicolumn{1}{c}{\ding{55}} & \multicolumn{1}{c}{\ding{55}} \\
    Statement terminators & \multicolumn{1}{c}{Newline} & \multicolumn{1}{c}{Semicolon} & \multicolumn{1}{c}{Semicolon (;)} & \multicolumn{1}{c}{Semicolon  (;)} & \multicolumn{1}{c}{Semicolon  (;)} & \multicolumn{1}{c}{Semicolon  (;)} \\
    Representation of assertion  & assert & assert & assert & Debug.Assert & assert & console.assert() \\
    \midrule
    \multicolumn{7}{c}{Object-oriented programming} \\
    \midrule
    Inheritance syntax & \multicolumn{1}{c}{class Dog (Animal)} & \multicolumn{1}{c}{class Dog: public Animal} & \multicolumn{1}{c}{class Dog extends Animal} & \multicolumn{1}{c}{class Dog: Animal} & \multicolumn{1}{c}{class Dog extends Animal} & \multicolumn{1}{c}{class Dog extends Animal} \\
    Syntax of private function & \multicolumn{1}{c}{def \_\_fname:} & \multicolumn{1}{c}{\makecell{private: \\ \qquad \qquad \ void fname}} & \multicolumn{1}{c}{private void fname} & \multicolumn{1}{c}{private void fname} & \multicolumn{1}{c}{private function fname} & \multicolumn{1}{c}{\#fname} \\
    Syntax of public function & def fname & \makecell{public: \\ \quad \quad \qquad void fname} & public void fname & public void fname & public function fname & fname \\
    Creating objects & 	obj = MyClass() & \makecell{obj = MyClass() or \\ obj = new MyClass()} & obj = new MyClass() & obj = new MyClass() & \$obj = new MyClass() & let obj = new MyClass() \\
    Constructor & 	def \_\_init\_\_(self, arr) & MyClass(std::vector<int> arr) & MyClass(int[] arr) & MyClass(int[] arr) & function \_\_construct(\$arr) & constructor(arr) \\
    Property initialization & 	self.arr = arr & this-\textgreater arr = arr & this.arr = arr & this.arr = arr & \$this-\textgreater arr = arr & this.arr = arr \\
    \bottomrule
    \end{tabular}%
    }
  \label{tab:tanslation_content}%
\end{table*}%

\subsection{Programming language selection}
\label{sec:PL_selection}
Based on the rankings of the four major programming language platforms, i.e., TIOBE~\footnote{\url{https://www.tiobe.com/tiobe-index/.}}, GitHut~\footnote{\url{https://madnight.github.io/githut/#/pull_requests/2024/1.}}, PYPL~\footnote{\url{https://pypl.github.io/PYPL.html.}}, and RedMonk~\footnote{\url{https://redmonk.com/sogrady/2024/03/08/language-rankings-1-24/.}}, and considering languages that support OOP, we selected the top six programming languages, i.e., C++, Java, C\#, PHP, Python, and JavaScript. We classified the six languages into three categories, i.e., high frequency, medium frequency, and low frequency, based on their popularity, as shown in Table~\ref{tab:programm_language}. 
Among these six languages, four (i.e., C++, C\#, PHP, and JavaScript) have never been used to evaluate the OOP capabilities of LLMs. Please refer to Appendix A in the supplementary materials for detailed classification criteria.

To address this gap, in this work, we consider translating the OOP benchmark into five other popular programming languages (i.e., C++, Java, C\#, PHP, and JavaScript), while retaining the Python programming language.

\subsection{Filtering of the OOP benchmark}
\label{sec:filtering}
It is well known that Python offers concise list comprehensions, making it highly intuitive and efficient for creating and manipulating lists. In contrast, other languages, such as Java, C++, and C\#, lack a fully equivalent built-in syntax. The OOP benchmark contains 431 problems, some of which require LLMs to return results in the form of nested lists, e.g., \emph{[[1, 2, 3], [4, 5, 6], [7, 8], [9, 10]]}. This format is not ideal for other programming languages like Java, C++, and C\#. Furthermore, Python supports multiple inheritance in OOP, while Java and C\# only support single inheritance and do not support multiple inheritance.

Therefore, a thorough review of the OOP benchmark is conducted, with 164 problems eliminated and 267 retained to enhance the accuracy of the translation process. By maintaining a balanced number of samples for each programming language, a more comprehensive understanding of LLMs’ performance across different languages is aimed for. After filtering, 267 problems are retained, including 593 class names, 177 private function names, 421 public function names, and 172 inheritance relationships. The detailed screening process can be found in Appendix B of the supplementary materials.

\subsection{Construction of the MultiOOP evaluation benchmark}
\label{sec:translation_OOP}
The construction of the MultiOOP evaluation benchmark mainly consists of three stages: translating requirement descriptions, unit tests, and matching function.

\subsubsection{Translation of requirement descriptions}
The problems in the OOP benchmark are requirements written in natural language. Therefore, during the translation process of these requirements, the translator only needs to convert the implementations from the target programming language of the problems into the other five programming languages.

\subsubsection{Translation of unit tests}
Our unit tests primarily consist of test cases and matching cases, as shown in Figure~\ref{fig:ovall_frame}. Test cases help evaluate the executability of code generated by LLMs, while matching cases assess whether the generated code meets the user's OOP concepts and feature requirements. To this end, our unit test translation process mainly focuses on two aspects: translating non-object-oriented programming content and translating object-oriented programming concepts and features, such as inheritance, encapsulation, and object creation.

\textbf{Translation of non-object-oriented programming content}. The translation of non-object-oriented programming content mainly includes the translation of variable types, statement terminators, and assertions. 
Python is a dynamically typed language, meaning variables in Python do not require explicit type declarations. The type of a variable is automatically inferred based on the assigned value. In contrast, languages like Java, C++, and C\# are statically typed, meaning variable types must be explicitly declared. Moreover, languages, e.g., Java, C++, and C\#, use semicolons to terminate statements, while Python uses newlines.
Therefore, we need to consider the translation of statement terminators. In addition, Python emphasizes simplicity and readability, which is why it uses the ``assert'' syntax. C\#, designed with the .NET ecosystem in mind, typically uses ``Debug.Assert'' for debugging and development. JavaScript, uses ``console.assert()'' for debugging in browsers or developer tools, reflecting JavaScript's flexibility and its focus on front-end development.

\begin{figure*}[!t]
    \centering
    \includegraphics[width=0.90\textwidth]{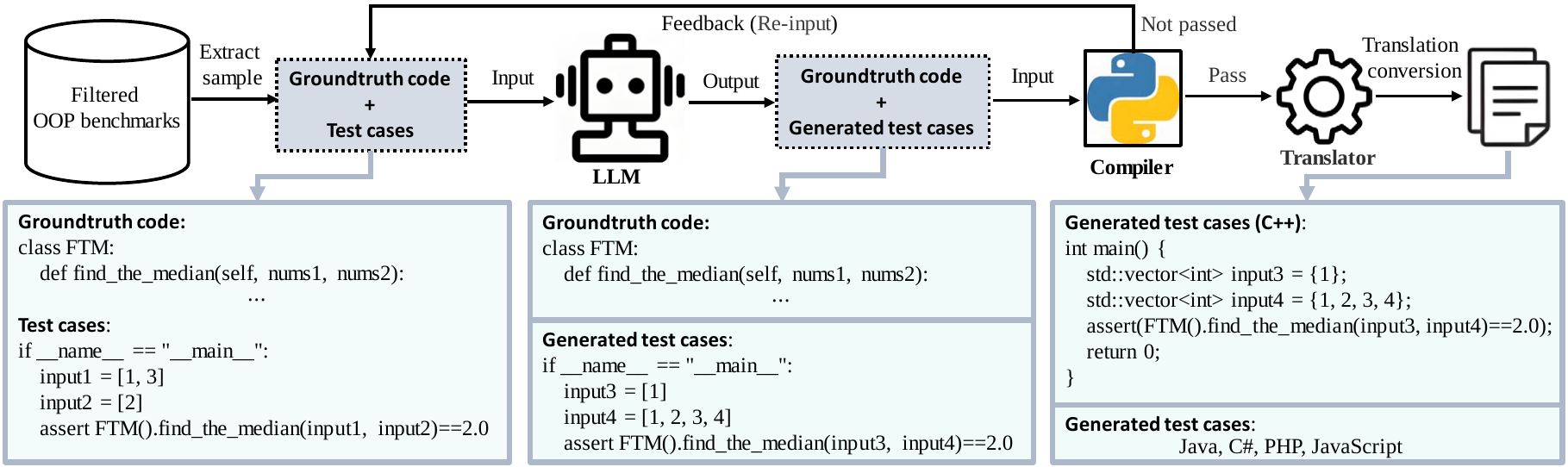}
    \caption{An automated framework for generating test cases.}
    \label{fig:case_frame}
\end{figure*}

\textbf{Translation of object-oriented programming concepts and features}. The translation of object-oriented programming involves the translation of concepts, e.g., representing inheritance, creating objects, private function, public function, etc. Python is a highly flexible and concise object-oriented programming language. It supports both single inheritance and multiple inheritance, and does not have enforced access control modifiers like private or protected. However, Python achieves encapsulation in object-oriented programming by convention, where member variables and methods that begin with a single underscore (i.e., ``\_'') or double underscore (i.e., ``\_\_'') are considered private or protected. In comparison, Java supports single inheritance and interface implementation but does not allow multiple inheritance. Nevertheless, Java enables similar functionality by implementing multiple interfaces and uses access control modifiers (i.e., private, protected, and public) to achieve encapsulation, with class member variables typically set to private. Additionally, other programming languages like C++, C\#, PHP, and JavaScript differ from Python in their object-oriented implementation. For example, C++ offers more powerful object-oriented features and greater flexibility. C++ not only supports single inheritance and multiple inheritance, but also uses access control modifiers to implement encapsulation, allowing member variables to be set to private, protected, or public. C++ also supports static polymorphism (e.g., function overloading) and dynamic polymorphism (e.g., virtual functions, inheritance, and polymorphism), with virtual functions being the key mechanism for achieving polymorphism in C++.

The content of the above unit tests translation process is summarized in Table~\ref{tab:tanslation_content}.

\subsubsection{Translation of matching function} The \textit{pass@$o$} metric~\cite{wang-etal-2024-oop} builds upon the \textit{pass@$k$} metric~\cite{chen2021evaluating} by further incorporating the matching between key points in natural language descriptions and key points in programming languages (e.g., class name, private function name, public function name). As a result, the translation of matching function becomes a crucial component for extending the \textit{pass@$o$} metric to a multi-language environment.
However, different programming languages use different terminology to express the same concepts, which makes cross-language translation and transformation more complex. For example, in Python, the for loop is iterator-based and typically iterates over iterable objects (e.g., lists or tuples) using the ``in'' keyword. In contrast, Java's for loop follows the C language style, with a syntax structure similar to that of C language, consisting of three parts: the initialization statement, the condition expression, and the update expression. 

Therefore, when performing matching function translation, special attention must be given to differences in loop statements and function return types to ensure that the \textit{pass@$o$} metric can be successfully extended and applied across multiple programming languages. 
For details of the translation process, please refer to Appendix C in the supplementary materials.


\subsection{Automated test case generation framework}
Due to the limited number of test cases in the OOP benchmark, it is insufficient for fully assessing the performance of LLMs in code generation. To this end, we have carefully designed an automated framework for generating test cases. This framework consists of three steps: generating new test cases, validating them, and translating them. Our designed automated framework is illustrated in Figure~\ref{fig:case_frame}.

\textbf{Generating new test cases}. We provide the LLM with the ground-truth Python code and several existing test cases, allowing it to reference them and generate new ones. The prompts used in this step can be found in Appendix D of the supplementary materials.

\textbf{Validating new test cases}. The newly generated test cases are incorporated into the Python code to verify their correctness through execution. We retain 18 passing test cases per sample from the filtered OOP benchmark during this process.

\textbf{Translating the test cases}. In this step, we apply the translator presented in section~\ref{sec:translation_OOP} to convert the newly generated Python test cases into equivalent ones in other programming languages (i.e., Java, C\#, PHP, and JavaScript).

\section{Experimental setup}
\label{sec:experimental_setup}
\subsection{Research questions}
In our experiments, we investigate the following research questions:

\noindent $\bullet$ \textbf{RQ1 (Overall Performance)}:
How well do LLMs perform on MultiOOP benchmark?

\noindent $\bullet$ \textbf{RQ2 (Performance of Different Programming Languages)}: Do LLMs exhibit significant performance differences in OOP tasks across different programming languages?

\noindent $\bullet$ \textbf{RQ3 (Performance on the \textit{Pass@$k$} and \textit{Pass@$o$} Metrics)}: 
Is the performance of LLMs the same on the \textit{pass@$k$} metric and the \textit{pass@$o$} metric? 

\noindent $\bullet$ \textbf{RQ4 (High Frequency vs. Low Frequency)}: Do LLMs perform better on MultiOOP benchmark in high-frequency programming languages than in low-frequency ones?


\noindent $\bullet$ \textbf{RQ5 (Prompting Selection)}: 
Can commonly used prompting methods such as zero-shot CoT prompting~\cite{wei2022chain} and few-shot prompting~\cite{brown2020language} enhance the performance of LLMs on MultiOOP benchmark? 

\noindent $\bullet$ \textbf{RQ6 (Bad Case Analysis)}: What are the common errors of LLMs in MultiOOP benchmark? 

\noindent $\bullet$ \textbf{RQ7 (The Impact of Test Cases)}: Why is using a larger number of test cases necessary to accurately reflect the performance of LLMs?

\subsection{Parameter settings}
In the experiment, we follow Roziere et al.~\cite{roziere2023code} by setting top\textit{-p} to 0.95 and adjusting the temperature to 0.1 and 0.8 during testing. We evaluate the MultiOOP benchmark using four NVIDIA A100 GPUs.

\subsection{Prompting design}
To evaluate the performance of the MultiOOP benchmark developed for this study under varying prompting conditions, we use the following three prompting methods:

\noindent $\bullet$ \textbf{Zero-shot prompting}: The user requirements are input directly into the LLM without the provision of any exemplar cases or guidance.


\noindent $\bullet$ \textbf{Zero-shot CoT prompting}:
The zero-shot prompting method is enhanced by adding the phrase ``let's think step by step'' to guide the LLM through sequential reasoning, resulting in more accurate outputs.

\noindent $\bullet$ \textbf{Few-shot prompting}: Zero-shot prompting is expanded on by providing several examples to help the LLM learn progressively and generate content that better meets expectations.

\subsection{Evaluated LLMs}
To objectively and fairly assess the MultiOOP performance of LLMs, we select nine general LLMs (e.g., Llama3-8b, Llama3.2-11b, Phi3-medium-4k-instruct, GPT-4o mini, etc) and five code-specialized LLMs (e.g., CodeLlama-7b, StarCoder, WizardCoder-15b-V1.0, etc).
Some of these models, such as StarCoder2, are pretrained using multiple programming languages. For detailed descriptions of these models, please refer to Appendix E in the supplementary materials.

\subsection{Metrics}
To objectively and fairly evaluate the performance of LLMs on the MultiOOP tasks, we use \textit{pass@$k$}~\cite{chen2021evaluating} and our multilingual extension \textit{pass@$o$} for assessment. 
The \textit{pass@$k$} metric is a widely used evaluation metric to assess whether code generated by LLMs passes predefined test cases. The \textit{pass@$o$} metric, based on this, imposes stricter requirements: the generated code must not only pass the test cases but also incorporate OOP constructs specified in the user's natural language description, such as class name, private function name, and public function name. Therefore, it is better suited for evaluating the structural complexity and fidelity of MultiOOP tasks.
Detailed descriptions of the \textit{pass@$k$} and \textit{pass@$o$} metrics can be found in Appendix F of the supplementary materials.

\begin{table*}[!t]
  \centering
  \caption{Performance of 14 mainstream large language models (LLMs) on MultiOOP benchmark. We also reported the differences in evaluation results between \textit{pass@$k$} and \textit{pass@$o$} (In the experiments, we set the temperature to 0.1, and all LLMs are evaluated in zero-shot prompting. Bold indicates the differences evaluated using the \textit{pass@$k$} and \textit{pass@$o$} metrics; Red indicates the best evaluation result; Underlined indicates the maximum disparities evaluated between \textit{pass@$k$} and \textit{pass@$o$} metrics; Gray indicates specialized code LLMs.). More results can be found in Appendix G of the supplementary materials.}
  \resizebox{1.0\linewidth}{!}{
    \begin{tabular}{l||ccccccccc||ccccccccc}
    \toprule
    \multicolumn{1}{l}{\multirow{2}[4]{*}{Model}} & \multicolumn{3}{c}{1} & \multicolumn{3}{c}{8} & \multicolumn{3}{c}{15} & \multicolumn{3}{c}{1} & \multicolumn{3}{c}{8} & \multicolumn{3}{c}{15} \\
\cmidrule{2-19}    \multicolumn{1}{c||}{} & $k$ & $o$ & $\boldsymbol{\Delta}\left(\downarrow\right)$     & $k$ & $o$ & $\boldsymbol{\Delta}\left(\downarrow\right)$     & $k$ & $o$ & \multicolumn{1}{c||}{$\boldsymbol{\Delta}\left(\downarrow\right)$} & $k$ & $o$ & $\boldsymbol{\Delta}\left(\downarrow\right)$     & $k$ & $o$ & W     & $k$ & $o$ & $\boldsymbol{\Delta}\left(\downarrow\right)$ \\
    \midrule
    \multicolumn{1}{c||}{} & \multicolumn{9}{c||}{Python (High)}                                            & \multicolumn{9}{c}{C++ (Medium)} \\
    \midrule
    Llama3-8b & 2.42  & 2.22  & \textbf{-0.20}  & 6.35  & 4.91  & \textbf{-1.44}  & 7.49  & 5.62  & \textbf{-1.87} 
    & 0.07  & 0.07  & \textbf{-0.00}  & 0.60  & 0.60  & \textbf{-0.00}  & 1.12  & 1.12  & \textbf{-0.00}  \\
    Llama3.1-8b & 1.32  & 1.17  & \textbf{-0.15}  & 5.49  & 4.64  & \textbf{-0.85}  & 7.49  & 6.37  & \textbf{-1.12} 
    & 0.00  & 0.00  & \textbf{-0.00}  & 0.04  & 0.01  & \textbf{-0.03}  & 0.11  & 0.03  & \textbf{-0.08}  \\
    Llama3.1-8b-Instruct & 13.18  & 11.64  & \textbf{-1.54}  & 22.77  & 19.33  & \textbf{-3.44}  & 24.72  & 20.97  & \textbf{-3.75}  
    & 1.60  & 1.55  & \textbf{-0.05}  & 4.23  & 4.03  & \textbf{-0.20}  & 5.24  & 4.87  & \textbf{-0.37}  \\
    Llama3.2-11b & 12.76  & 6.72  & \textbf{-6.04}  & 21.86  & 21.61  & \textbf{-0.25}  & 23.60  & 22.72  & \textbf{-0.88}  
    & 1.62  & 1.60  & \textbf{-0.02}  & 3.91  & 3.71  & \textbf{-0.20}  & 4.49  & 4.12  & \textbf{-0.37}  \\
    Phi3-medium-4k-instruct & 29.54  & 28.41  & \textbf{-1.13}  & 42.30  & 39.36  & \textbf{-2.94}  & 45.32  & 42.70  & \textbf{-2.62}  
    & 5.39  & 3.80  & \textbf{-1.59}  & 11.35  & 8.95  & \textbf{-2.40}  & 13.11  & 10.49  & \textbf{-2.62}  \\
    Qwen2.5-7b-Instruct & 24.94  & 16.70  & \textbf{-8.24}  & 48.09  & 31.32  & \underline{\textbf{-16.77}}  & 52.06  & 34.08  & \underline{\textbf{-17.98}}  
    & 4.52  & 0.62  & \textbf{-3.90}  & 10.39  & 2.60  & \textbf{-7.79}  & 11.99  & 3.37  & \textbf{-8.62}  \\
    Qwen2.5-14b-Instruct & 34.23  & 25.77  & \textbf{-8.46}  & 59.38  & 49.30  & \textbf{-10.08}  & 63.67  & 53.93  & \textbf{-9.74}  
    & 10.61  & 5.47  & \textbf{-5.14}  & 20.71  & 11.72  & \underline{\textbf{-8.99}}  & 23.22  & 13.48  & \underline{\textbf{-9.74}}  \\
    Vicuna-13b-v1.5 & 2.17  & 1.65  & \textbf{-0.52}  & 5.55  & 5.00  & \textbf{-0.55}  & 6.74  & 5.99  & \textbf{-0.75}  
    & 0.02  & 0.02  & \textbf{-0.00}  & 0.20  & 0.20  & \textbf{-0.00}  & 0.37  & 0.37  & \textbf{-0.00}  \\
    GPT-4o mini & \textcolor{red}{\textbf{58.85}}  & \textcolor{red}{\textbf{48.06}}  & \textbf{-10.79}  & \textcolor{red}{\textbf{68.00}}  & \textcolor{red}{\textbf{59.29}}  & \textbf{-8.71}  & \textcolor{red}{\textbf{69.29}}  & \textcolor{red}{\textbf{61.42}}  & \textbf{-7.87}  
    & \textcolor{red}{\textbf{23.05}}  & \textcolor{red}{\textbf{15.26}}  & \underline{\textbf{-7.79}}  & \textcolor{red}{\textbf{27.51}}  & \textcolor{red}{\textbf{19.46}}  & \textbf{-8.05}  & \textcolor{red}{\textbf{28.46}}  & \textcolor{red}{\textbf{20.60}}  & \textbf{-7.86}  \\
    \rowcolor[rgb]{ .906,  .902,  .902} CodeLlama-7b & 3.55  & 2.02  & \textbf{-1.53}  & 11.25  & 7.10  & \textbf{-4.15}  & 13.86  & 8.99  & \textbf{-4.87}  
    & 0.60  & 0.55  & \textbf{-0.05}  & 2.14  & 1.74  & \textbf{-0.40}  & 3.00  & 2.25  & \textbf{-0.75}  \\
    \rowcolor[rgb]{ .906,  .902,  .902} CodeLlama-13b & 9.49  & 6.87  & \textbf{-2.62}  & 20.71  & 13.72  & \textbf{-6.99}  & 23.97  & 15.36  & \textbf{-8.61}  
    & 1.00  & 0.92  & \textbf{-0.08}  & 2.32  & 1.98  & \textbf{-0.34}  & 2.62  & 2.25  & \textbf{-0.37}  \\
    \rowcolor[rgb]{ .906,  .902,  .902} StarCoder & 4.00  & 1.40  & \textbf{-2.60}  & 8.96  & 5.29  & \textbf{-3.67}  & 10.49  & 6.74  & \textbf{-3.75}
    & 2.02  & 0.10  & \textbf{-1.92}  & 4.16  & 0.70  & \textbf{-3.46}  & 4.62  & 1.12  & \textbf{-3.50}  \\
    \rowcolor[rgb]{ .906,  .902,  .902} StarCoder2 & 26.77  & 15.91  & \underline{\textbf{-10.86}}  & 35.54  & 21.28  & \textbf{-14.26}  & 37.83  & 22.85  & \textbf{-14.98}  
    & 2.02  & 2.02  & \textbf{-0.00}  & 4.16  & 4.16  & \textbf{-0.00}  & 4.87  & 4.87  & \textbf{-0.00}  \\
    \rowcolor[rgb]{ .906,  .902,  .902} WizardCoder-15b-V1.0 & 9.04  & 6.39  & \textbf{-2.65}  & 16.63  & 12.44  & \textbf{-4.19}  & 18.35  & 13.86  & \textbf{-4.49}  
    & 1.60  & 1.27  &\textbf{-0.33}  & 2.74  & 2.00  & \textbf{-0.74}  & 3.00  & 2.25  & \textbf{-0.75}  \\
    \midrule
    \multicolumn{1}{c||}{} & \multicolumn{9}{c||}{Java (High)}                                            & \multicolumn{9}{c}{C\# (Low)} \\
    \midrule
    Llama3-8b & 0.02  & 0.00  & \textbf{-0.02}  & 0.20  & 0.04  & \textbf{-0.16}  & 0.37  & 0.06  & \textbf{-0.31} 
    & 0.25  & 0.22  & \textbf{-0.03}  & 0.67  & 0.65  & \textbf{-0.02}  & 0.75  & 0.75  & \textbf{-0.00}  \\
    Llama3.1-8b & 0.02  & 0.02  & \textbf{-0.00}  & 0.20  & 0.20  & \textbf{-0.00}  & 0.37  & 0.37  & \textbf{-0.00} 
    & 1.25  & 1.17  & \textbf{-0.08}  & 7.01  & 4.64  & \textbf{-2.37}  & 7.81  & 6.37  & \textbf{-1.44}  \\
    Llama3.1-8b-Instruct & 0.50  & 0.47  & \textbf{-0.03}  & 0.95  & 0.75  & \textbf{-0.20}  & 1.12  & 0.78  & \textbf{-0.34}  
    & 0.12  & 0.12  & \textbf{-0.00}  & 0.65  & 0.65  & \textbf{-0.00}  & 0.75  & 0.75  & \textbf{-0.00}  \\
    Llama3.2-11b & 0.42  & 0.40  & \textbf{-0.02}  & 1.28  & 1.09  & \textbf{-0.19}  & 1.50  & 1.12  & \textbf{-0.38} 
    & 0.37  & 0.30  & \textbf{-0.07}  & 1.22  & 0.92  & \textbf{-0.30}  & 1.50  & 1.12  & \textbf{-0.38}  \\
    Phi3-medium-4k-instruct & 5.79  & 3.17  & \textbf{-2.62}  & 9.31  & 4.99  & \textbf{-4.32}  & 10.11  & 5.62  & \textbf{-4.49}  
    & 3.32  & 2.17  & \textbf{-1.15}  & 7.66  & 5.19  & \textbf{-2.47}  & 8.99  & 6.37  & \textbf{-2.62}  \\
    Qwen2.5-7b-Instruct & 3.80  & 0.47  & \textbf{-3.33}  & 8.69  & 1.37  & \textbf{-7.32}  & 10.86  & 2.25  & \textbf{-8.61}  
    & 0.35  & 0.12  & \textbf{-0.23}  & 1.54  & 0.56  & \textbf{-0.98}  & 2.25  & 0.75  & \textbf{-1.50}  \\
    Qwen2.5-14b-Instruct & 7.39  & 2.05  & \textbf{-5.34}  & 14.41  & 6.01  & \textbf{-8.40}  & 15.73  & 7.12  & \textbf{-8.61}  
    & 0.50  & 0.25  & \textbf{-0.25}  & 2.85  & 1.64  & \textbf{-1.21}  & 4.12  & 2.62  & \textbf{-1.50}  \\
    Vicuna-13b-v1.5 & 0.07  & 0.00  & \textbf{-0.07}  & 0.50  & 0.05  & \textbf{-0.45}  & 0.75  & 0.16  & \textbf{-0.59}  
    & 0.15  & 0.07  & \textbf{-0.08}  & 0.85  & 0.35  & \textbf{-0.50}  & 1.12  & 0.38  & \textbf{-0.74}  \\
    GPT-4o mini & \textcolor{red}{\textbf{16.98}}  & 0.77  & \underline{\textbf{-16.21}}  & \textcolor{red}{\textbf{20.51}}  & 1.29  & \underline{\textbf{-19.22}}  & \textcolor{red}{\textbf{21.34}}  & 1.50  & \underline{\textbf{-19.84}}  
    & \textcolor{red}{\textbf{21.72}}  & \textcolor{red}{\textbf{12.61}}  & \underline{\textbf{-9.11}}  & \textcolor{red}{\textbf{25.23}}  & \textcolor{red}{\textbf{15.67}}  & \underline{\textbf{-9.56}}  & \textcolor{red}{\textbf{25.84}}  & \textcolor{red}{\textbf{16.48}}  & \underline{\textbf{-9.36}}  \\
    \rowcolor[rgb]{ .906,  .902,  .902} CodeLlama-7b & 0.45  & 0.17  & \textbf{-0.28}  & 2.51  & 0.96  & \textbf{-1.55}  & 3.37  & 1.50  & \textbf{-1.87} 
    & 0.75  & 0.47  & \textbf{-0.28}  & 3.93  & 2.47  & \textbf{-1.46}  & 5.62  & 3.37  & \textbf{-2.25}  \\
    \rowcolor[rgb]{ .906,  .902,  .902} CodeLlama-13b & 1.30  & 0.07  & \textbf{-1.23}  & 3.64  & 0.50  & \textbf{-3.14}  & 4.12  & 0.75  & \textbf{-3.37}
    & 0.90  & 0.40  & \textbf{-0.50}  & 2.52  & 1.47  & \textbf{-1.05}  & 3.37  & 2.25  & \textbf{-1.12}  \\
    \rowcolor[rgb]{ .906,  .902,  .902} StarCoder & 0.02  & 0.01  & \textbf{-0.01}  & 0.20  & 0.09  & \textbf{-0.11}  & 0.37  & 0.16  & \textbf{-0.21} 
    & 0.12  & 0.12  & \textbf{-0.00}  & 0.65  & 0.65  & \textbf{-0.00}  & 0.75  & 0.75  & \textbf{-0.00}  \\
    \rowcolor[rgb]{ .906,  .902,  .902} StarCoder2 & 5.29  & \textcolor{red}{\textbf{5.29}}  & \textbf{-0.00}  & 8.22  & \textcolor{red}{\textbf{8.22}}  & \textbf{-0.00}  & 8.99  & \textcolor{red}{\textbf{8.99}}  & \textbf{-0.00} 
    & 1.67  & 1.37  & \textbf{-0.30}  & 4.42  & 3.92  & \textbf{-0.50}  & 5.24  & 4.46  & \textbf{-0.78}  \\
    \rowcolor[rgb]{ .906,  .902,  .902} WizardCoder-15b-V1.0 & 2.30  & 0.77  & \textbf{-1.53}  & 3.42  & 1.15  & \textbf{-2.27}  & 4.12  & 1.50  & \textbf{-2.62}  
    & 1.07  & 0.92  & \textbf{-0.15}  & 3.11  & 2.37  & \textbf{-0.74}  & 3.75  & 3.00  & \textbf{-0.75}  \\
    \midrule
    \multicolumn{1}{c||}{} & \multicolumn{9}{c||}{PHP (Low)}                                            & \multicolumn{9}{c}{JavaScript (Medium)} \\
    \midrule
    Llama3-8b & 0.12  & 0.02  & \textbf{-0.10}  & 0.90  & 0.20  & \textbf{-0.70}  & 1.50  & 0.37  & \textbf{-1.13} 
    & 1.47  & 1.35  & \textbf{-0.12}  & 5.11  & 4.21  & \textbf{-0.90}  & 6.74  & 5.24  & \textbf{-1.50}  \\
    Llama3.1-8b & 0.82  & 0.45  & \textbf{-0.37}  & 5.20  & 2.99  & \textbf{-2.21}  & 8.24  & 4.87  & \textbf{-3.37} 
    & 1.27  & 1.17  & \textbf{-0.10}  & 2.89  & 2.34  & \textbf{-0.55}  & 3.37  & 2.62  & \textbf{-0.75}  \\
    Llama3.1-8b-Instruct & 9.79  & 0.75  & \textbf{-9.04}  & 17.15  & 1.79  & \textbf{-15.36}  & 19.48  & 1.87  & \textbf{-17.61}  
    & 6.57  & 5.69  & \textbf{-0.88}  & 13.95  & 11.74  & \textbf{-2.24}  & 16.48  & 14.23  & \textbf{-2.25}  \\
    Llama3.2-11b & 11.14  & 1.05  & \textbf{-10.09}  & 20.24  & 2.05  & \textbf{-18.19}  & 22.85  & 2.25  & \textbf{-20.60}  
    & 4.99  & 4.64  & \textbf{-0.35}  & 9.17  & 8.18  & \textbf{-0.99}  & 10.11  & 8.99  & \textbf{-1.12}  \\
    Phi3-medium-4k-instruct & 19.90  & 2.12  & \textbf{-17.78}  & 32.94  & 4.49  & \textbf{-28.45}  & 35.58  & 5.62  & \textbf{-29.96}  
    & 9.84  & 9.84  & \textbf{-0.00}  & 19.29  & 19.08  & \textbf{-0.21}  & 21.72  & 20.97  & \textbf{-0.75}  \\
    Qwen2.5-7b-Instruct & 17.18  & 0.92  & \textbf{-16.26}  & 39.05  & 4.05  & \textbf{-35.00}  & 46.44  & 5.62  & \textbf{-40.82}  & 17.20  & 7.84  & \textbf{-9.36}  & 36.15  & 19.43  & \textbf{-16.72}  & 40.82  & 22.85  & \textbf{-17.97}  \\
    Qwen2.5-14b-Instruct & 30.46  & 2.35  & \textbf{-28.11}  & 56.82  & 6.19  & \textbf{-50.63}  & 62.92  & 7.49  & \textbf{-55.43}  
    & 23.20  & 13.56  & \textbf{-9.64}  & 51.83  & 33.53  & \textbf{-18.30}  & 57.30  & 39.33  & \textbf{-17.97}  \\
    Vicuna-13b-v1.5 & 0.05  & 0.00  & \textbf{-0.05}  & 0.40  & 0.02  & \textbf{-0.38}  & 0.75  & 0.10  & \textbf{-0.65}  
    & 0.50  & 0.25  & \textbf{-0.25}  & 1.49  & 0.57  & \textbf{-0.92}  & 1.87  & 0.75  & \textbf{-1.12}  \\
    GPT-4o mini & \textcolor{red}{\textbf{57.63}}  & \textcolor{red}{\textbf{4.42}}  & \underline{\textbf{-53.21}}  & \textcolor{red}{\textbf{69.06}}  & 5.57  & \underline{\textbf{-63.49}}  & \textcolor{red}{\textbf{71.54}}  & 5.99  & \underline{\textbf{-65.55}}  
    & \textcolor{red}{\textbf{52.36}}  & \textcolor{red}{\textbf{30.44}}  & \underline{\textbf{-21.92}}  & \textcolor{red}{\textbf{60.15}}  & \textcolor{red}{\textbf{39.32}}  & \underline{\textbf{-20.83}}  & \textcolor{red}{\textbf{62.17}}  & \textcolor{red}{\textbf{41.57}}  & \underline{\textbf{-20.60}}  \\
    \rowcolor[rgb]{ .906,  .902,  .902} CodeLlama-7b & 2.15  & 0.42  & \textbf{-1.73}  & 8.33  & 2.23  & \textbf{-6.10}  & 10.49  & 3.00  & \textbf{-7.49}
    & 3.30  & 1.72  & \textbf{-1.58}  & 11.41  & 6.22  & \textbf{-5.19}  & 14.23  & 7.49  & \textbf{-6.74}  \\
    \rowcolor[rgb]{ .906,  .902,  .902} CodeLlama-13b & 2.87  & 0.12  & \textbf{-2.75}  & 7.82  & 0.56  & \textbf{-7.26}  & 9.36  & 0.75  & \textbf{-8.61}
    & 1.65  & 0.57  & \textbf{-1.08}  & 5.85  & 1.91  & \textbf{-3.94}  & 7.49  & 2.62  & \textbf{-4.87}  \\
    \rowcolor[rgb]{ .906,  .902,  .902} StarCoder & 0.02  & 0.00  & \textbf{-0.02}  & 0.20  & 0.03  & \textbf{-0.17}  & 0.37  & 0.09  & \textbf{-0.28} 
    & 0.05  & 0.00  & \textbf{-0.05}  & 0.14  & 0.03  & \textbf{-0.11}  & 0.28  & 0.07  & \textbf{-0.21}  \\
    \rowcolor[rgb]{ .906,  .902,  .902} StarCoder2 & 3.90  & 3.90  & \textbf{-0.00}  & 9.38  & \textcolor{red}{\textbf{9.38}}  & \textbf{-0.00}  & 11.61  & \textcolor{red}{\textbf{11.61}}  & \textbf{-0.00}  
    & 17.45  & 16.93  & \textbf{-0.52}  & 30.04  & 28.38  & \textbf{-1.66}  & 32.58  & 31.09  & \textbf{-1.49}  \\
    \rowcolor[rgb]{ .906,  .902,  .902} WizardCoder-15b-V1.0 & 5.64  & 2.10  & \textbf{-3.54}  & 10.98  & 3.29  & \textbf{-7.69}  & 12.36  & 3.75  & \textbf{-8.61}  
    & 8.31  & 3.80  & \textbf{-4.51}  & 11.14  & 5.36  & \textbf{-5.78}  & 11.99  & 5.99  & \textbf{-6.00}  \\
    \bottomrule
    \end{tabular}%
    }
  \label{tab:result_01}%
\end{table*}%

\section{RESULTS}
Based on the experimental setup in section~\ref{sec:experimental_setup}, the performance of 14 mainstream LLMs on MultiOOP benchmark is evaluated, with specific results presented in Table~\ref{tab:result_01}. 


\subsection{RQ1: Overall Performance}
The evaluation results in Table~\ref{tab:result_01} clearly show that most LLMs perform relatively poorly in MultiOOP tasks. For instance, Llama3.1-8b achieves \textit{pass@$1$} scores of $1.17$, $0.00$, $0.02$, $1.17$, $0.45$, and $1.17$ in programming languages such as Python, C++, Java, C\#, PHP, and JavaScript, respectively. 
In contrast, only a very small number of LLMs achieve high scores in certain programming languages. For example, GPT-4o mini has \textit{pass@$1$} scores of $48.06$, $15.26$, and $30.44$ in Python, C++, and JavaScript, respectively. These findings underscore the overall limitations of current LLMs in handling OOP tasks and highlight considerable disparities in their capabilities. 
At the same time, this observation indirectly indicates that a small number of high-performing LLMs may have adopted more effective strategies in the design of their architectures or in their training processes, thereby achieving better performance on MultiOOP tasks.

\noindent \textbf{Answer to RQ1}:
In MultiOOP benchmark, the majority of LLMs (e.g., Llama3-8b, Llama3.1-8b, Vicuna-13b-v1.5, CodeLlama-7b, and CodeLlama-13b) show lower performance, with only a few (e.g., GPT-4o mini) capable of achieving higher performance in certain programming languages.

\begin{table*}[!t]
  \centering
  \caption{A performance comparison of LLMs in generating MultiOOP tasks under zero-shot CoT, and few-shot settings. The bold indicates the performance difference between zero-shot CoT, few-shot, and zero-shot settings.}
  \resizebox{1.0\linewidth}{!}{
    \begin{tabular}{ccccccccccccccccccc}
    \toprule
    Model & \multicolumn{6}{c}{Qwen2.5-14b-Instruct}     & \multicolumn{6}{c}{StarCoder2} & \multicolumn{6}{c}{GPT-4o mini}\\
    \midrule
    \textit{Pass@$o$} & 1     & $\boldsymbol{\Delta}\left(\uparrow\right)$     & 8     & $\boldsymbol{\Delta}\left(\uparrow\right)$     & 15    & $\boldsymbol{\Delta}\left(\uparrow\right)$     & 1     & $\boldsymbol{\Delta}\left(\uparrow\right)$     & 8     & $\boldsymbol{\Delta}\left(\uparrow\right)$     & 15    & $\boldsymbol{\Delta}\left(\uparrow\right)$ & 1     & $\boldsymbol{\Delta}\left(\uparrow\right)$     & 8     & $\boldsymbol{\Delta}\left(\uparrow\right)$     & 15    & $\boldsymbol{\Delta}\left(\uparrow\right)$\\
    \midrule
    \multicolumn{19}{c}{Python (High)} \\
    \midrule
    Zero-shot CoT  & 30.44  & \textbf{+4.67}  & 52.01  & \textbf{+2.71}  & 55.37  & \textbf{+1.80}  
    & 4.44  & \textbf{-11.47}  & 10.80  & \textbf{-10.48}  & 11.99  & \textbf{-10.86}  
    & 38.40  & \textbf{-9.66}  & 55.74  & \textbf{-3.55} & 59.18  & \textbf{-2.24}\\
    Few-shot & 50.09  & \textbf{+24.32}  & 58.21  & \textbf{+8.91}  & 60.67  & \textbf{+6.74}  
    & 33.71  & \textbf{+17.80}  & 38.11  & \textbf{+16.83}  & 39.33  & \textbf{+16.48}  
    & 58.40  & \textbf{+10.34}  & 65.74  & \textbf{+6.45}  & 67.18  & \textbf{+5.76}\\
    \midrule
    \multicolumn{19}{c}{C++ (Medium)} \\
    \midrule
    Zero-shot CoT   & 7.22  & \textbf{+1.75}  & 11.94  & \textbf{+0.22}  & 13.11  & \textbf{-0.37}  
    & 0.09  & \textbf{-1.93}  & 0.87  & \textbf{-3.29}  & 1.03 &\textbf{-3.84} 
    & 6.19  & \textbf{-9.07}  & 12.05  & \textbf{-7.89}  & 13.86  & \textbf{-8.99}\\
    Few-shot & 11.35  & \textbf{+5.88}  & 14.08  & \textbf{+2.36}  & 15.24  & \textbf{+1.76}  
    & 2.22  & \textbf{+0.20}  & 4.93  & \textbf{+0.77}  &5.12  & \textbf{+0.25}  
    & 15.73  & \textbf{+0.47}  & 29.41  & \textbf{+9.95}  & 30.62  & \textbf{+10.02}\\
    \midrule
    \multicolumn{19}{c}{Java (High)} \\
    \midrule
    Zero-shot CoT   & 2.10  & \textbf{+0.05}  & 5.09  & \textbf{-0.92}  & 6.37  & \textbf{-0.75}  
    & 0.10  & \textbf{-5.19}  & 0.70  & \textbf{-7.52}  & 1.12  & \textbf{-7.87}  
    & 6.72  & \textbf{+5.95}  & 11.84  & \textbf{+10.55}  & 12.64  & \textbf{+11.14}\\
    Few-shot & 10.62  & \textbf{+8.57}  & 18.11  & \textbf{+12.10}  & 19.13  & \textbf{+12.01}  
    & 14.61  & \textbf{+9.32}  & 22.37  & \textbf{+14.15}  & 24.16  & \textbf{+15.17}  
    & 8.91  & \textbf{+8.41}  & 15.22  & \textbf{+13.93}  & 17.64  & \textbf{+16.41}\\
    \midrule
    \multicolumn{19}{c}{C\# (Low)} \\
    \midrule
    Zero-shot CoT   & 2.66  & \textbf{+2.41}  & 6.33  & \textbf{+4.69}  & 7.12  & \textbf{+4.50}  
    & 1.07  & \textbf{-0.30}  & 2.50  & \textbf{-1.42}  & 2.62  & \textbf{-1.84}  
    & 5.57  & \textbf{-7.04}  & 10.37  & \textbf{-5.30}  & 11.99  & \textbf{-4.49}\\
    Few-shot & 1.07  & \textbf{+0.82}  & 2.50  & \textbf{+0.86}  & 2.62  & \textbf{+0.00}  
    & 2.66  & \textbf{+1.29}  & 6.33  & \textbf{+2.41}  & 7.12  & \textbf{+2.66}  
    & 12.84  & \textbf{+0.23}  & 16.28  & \textbf{+0.61}  & 17.04  & \textbf{+0.56}\\
    \midrule
    \multicolumn{19}{c}{PHP (Low)} \\
    \midrule
    Zero-shot CoT   & 5.37  & \textbf{+3.02}  & 12.02  & \textbf{+5.83}  & 13.97  & \textbf{+6.48}  
    & 4.15  & \textbf{+0.25}  & 10.08  & \textbf{+0.70}  & 11.87  & \textbf{+0.26}  
    & 11.58  & \textbf{+7.16}  & 23.65  & \textbf{+18.08}  & 25.79  & \textbf{+19.80}\\
    Few-shot & 5.22  & \textbf{+2.87}  & 5.54  & \textbf{-0.65}  & 5.62  & \textbf{-1.87}  
    & 4.62  & \textbf{+0.72}  & 10.05  & \textbf{+0.67}  & 11.12  & \textbf{-0.49}  
    & 5.24  & \textbf{+0.82}  & 7.43  & \textbf{+1.86}  & 8.24  & \textbf{+2.25}\\
    \midrule
    \multicolumn{19}{c}{JavaScript (Medium)} \\
    \midrule
    Zero-shot CoT   & 17.50  & \textbf{+3.94}  & 32.59  & \textbf{-0.49}  & 37.08  & \textbf{-2.25}  
    & 1.90  & \textbf{-15.03}  & 5.38  & \textbf{-23.00}  & 6.74  & \textbf{-24.35}  
    & 15.21  & \textbf{-15.23}  & 25.97  & \textbf{-13.25}  & 27.04  & \textbf{-14.57}\\
    Few-shot & 20.15  & \textbf{+6.59}  & 25.22  & \textbf{-8.31}  & 25.84  & \textbf{-13.49} 
    & 17.78  & \textbf{+0.85}  & 30.55  & \textbf{+2.17}  & 32.60  & \textbf{+1.51}  
    & 32.97  & \textbf{+2.53}  & 43.16  & \textbf{+3.84}  & 45.09  & \textbf{+3.52}\\
    \bottomrule
    \end{tabular}%
    }
  \label{tab:prompt_selection}%
\end{table*}%

\begin{figure*}[!t]
    \centering
    \begin{subfigure}{0.47\textwidth}
        \centering
        \includegraphics[width=\linewidth]{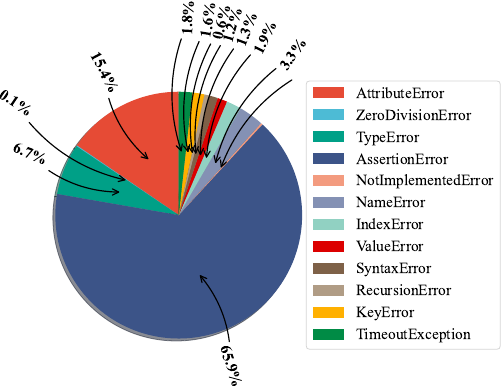}
        \caption{Python}
    \end{subfigure}
    \hspace{0.005cm}
    \begin{subfigure}{0.47\textwidth}
        \centering
        \includegraphics[width=\linewidth]{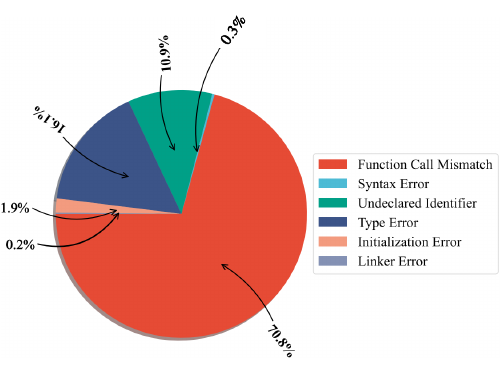}
        \caption{C++}
    \end{subfigure}
    \caption{Error distribution of Python and C++ programming languages.}
    \label{fig:error_distribution}%
\end{figure*}

\begin{figure*}[!t]
    \centering
    \begin{subfigure}{0.49\textwidth}
        \centering
        \includegraphics[width=\linewidth]{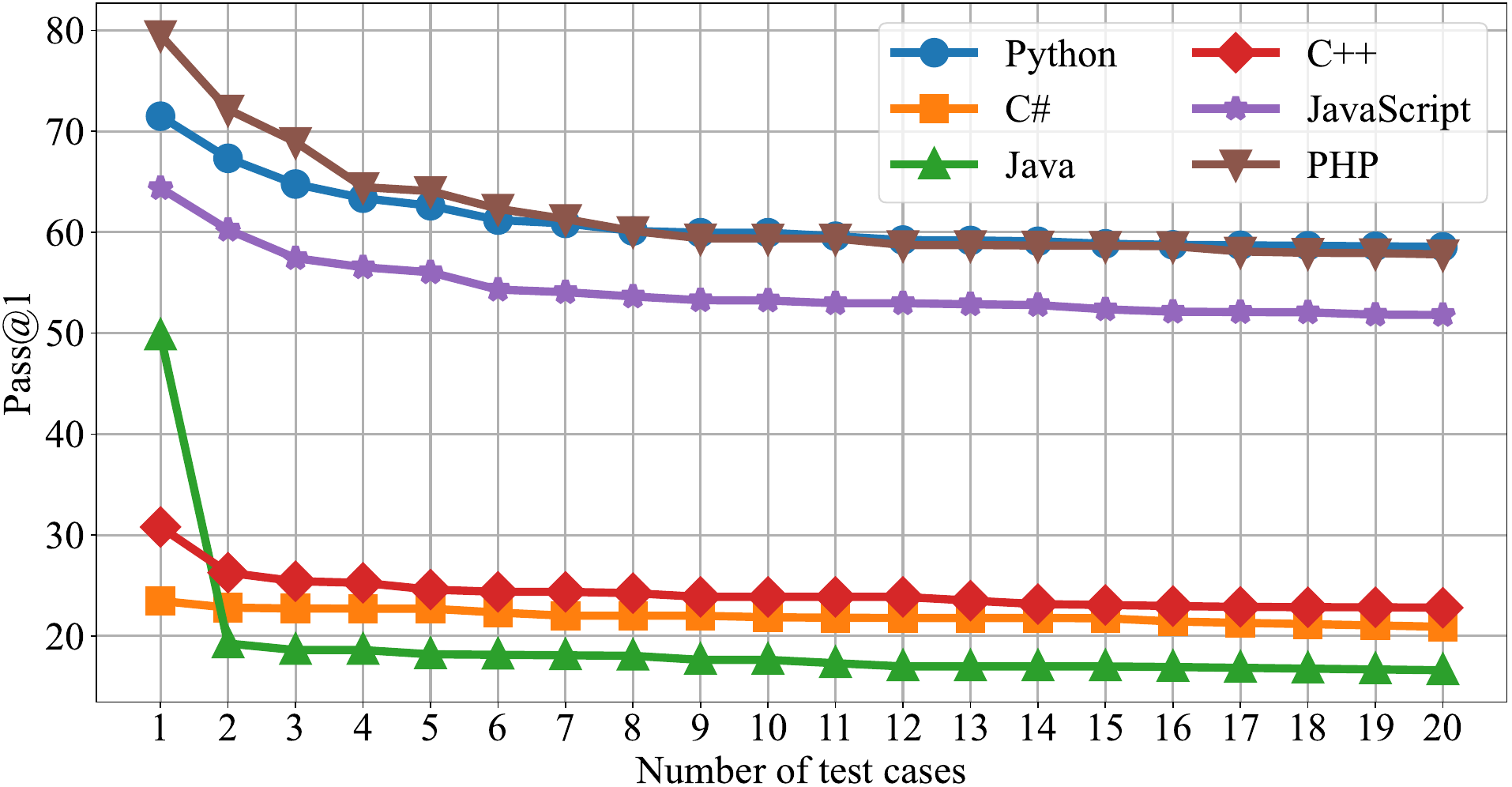}
        \caption{GPT-4o mini}
    \end{subfigure}
    \hspace{0.005cm}
    \begin{subfigure}{0.49\textwidth}
        \centering
        \includegraphics[width=\linewidth]{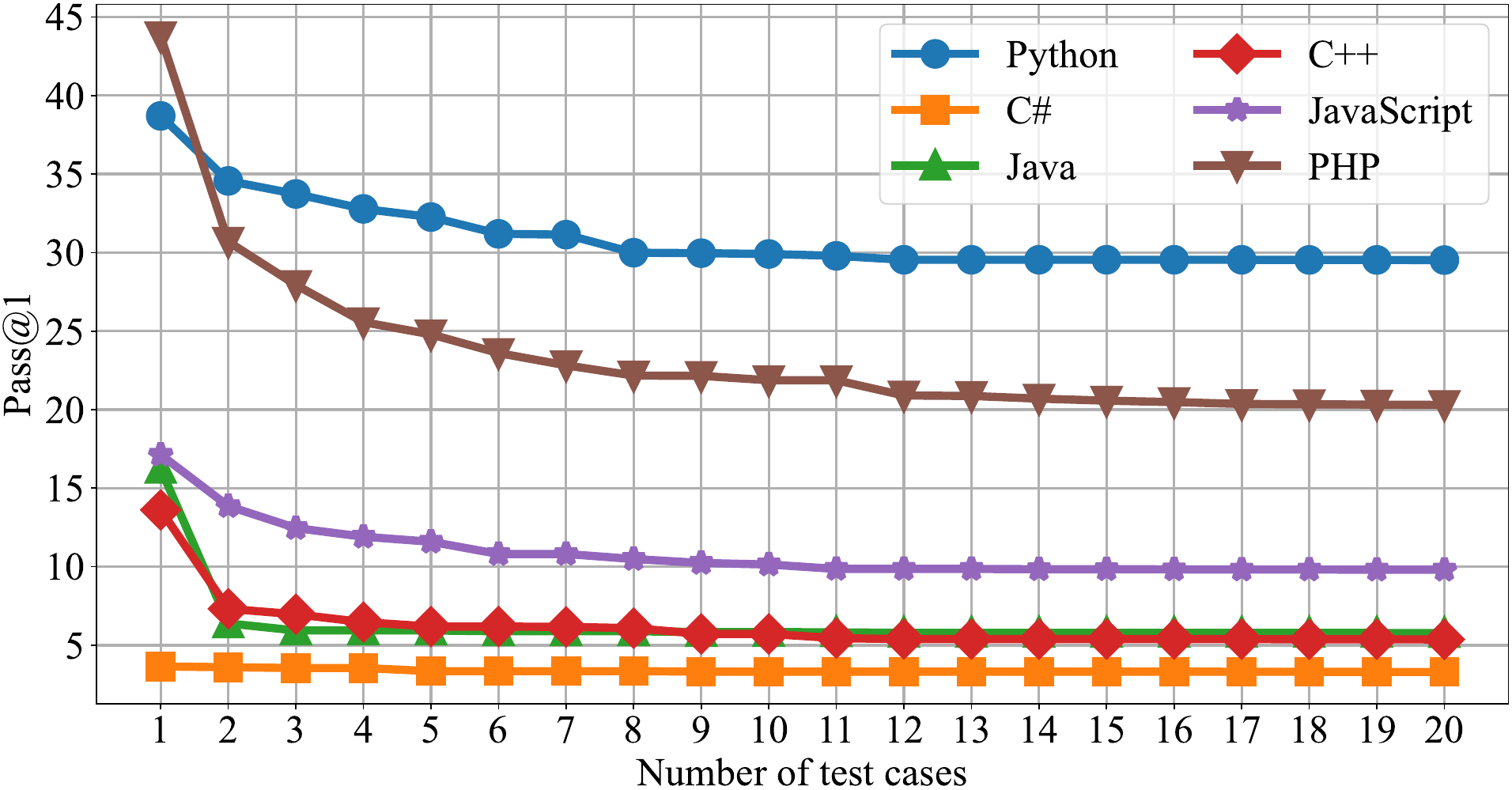}
        \caption{Phi3-medium-4k-instruct}
    \end{subfigure}
    \caption{Scores of LLMs (i.e., GPT-4o mini and Phi3-medium-4k-instruct) with an increasing number of test cases. Here, we use \textit{pass@$k$} as the evaluation metric. Further details of the test cases are provided in Appendix I of the supplementary materials.}
    \label{tab:GPT-4o mini_performance_trend}%
\end{figure*}

\subsection{RQ2: Performance of Different Programming Languages}
As shown in Table~\ref{tab:result_01}, LLMs exhibit notable disparities in performance on OOP tasks across programming languages. For example, GPT-4o mini achieves strong results in Python (\textit{pass@$1$}: $48.06$; \textit{pass@$8$}: $59.29$; \textit{pass@$15$}: $61.42$), but its performance drops sharply in C\# (\textit{pass@$1$}: $12.61$; \textit{pass@$1$}: $15.67$; \textit{pass@$15$}: $16.48$). Similar cross language fluctuations are observed in LLMs such as Phi-3 Medium-4k-Instruct, Qwen2.5-7B-Instruct, Qwen2.5-14B-Instruct, and StarCoder2. In comparison, LLMs like Llama3-8B show fairly stable but consistently low performance across different languages.
For instance, its \textit{pass@$1$} scores range from $0.00$ (Java) to $2.22$ (Python), and \textit{pass@$15$} from $0.06$ to $5.62$. Other LLMs including Llama3.1-8b, Vicuna-13b-v1.5, and CodeLlama-7b show similar stability. These findings indicate: 1) most LLMs exhibit significant performance variations in OOP tasks across different programming languages, with limited cross language generalization; 2) LLMs like GPT-4o mini perform well in some languages but experience substantial performance drops in others, suggesting a bias toward mainstream languages; 3) LLMs with consistent performance (e.g., Llama3-8b, CodeLlama-7b), though less accurate, show smaller fluctuations across languages, potentially having learned more universal structural features.

\noindent\textbf{Answer to RQ2}:
Some LLMs (such as Llama3-8b, Vicuna-13B-v1.5, and CodeLlama-7b) show relatively small performance differences across OOP tasks in different programming languages, while other LLMs (such as GPT-4o mini, Phi-3 Medium-4k-Instruct, Qwen2.5-7B-Instruct, and Qwen2.5-14B-Instruct) exhibit larger performance variations on these tasks across different programming languages.

\subsection{RQ3: Performance on the \textit{Pass@$k$} and \textit{Pass@$o$} metrics}
As shown in Table~\ref{tab:result_01}, for the Python language, all $14$ popular LLMs (e.g., Llama3-8b, GPT-4o mini) have lower \textit{pass@$o$} scores than \textit{pass@$k$}. Notably, StarCoder2 shows the largest gap on \textit{pass@$1$} (10.86), while Qwen2.5-7b-Instruct shows the most significant differences on \textit{pass@$8$} and \textit{pass@$15$} (16.77 and 17.98, respectively). In Java, only Llama3.1-8b and StarCoder2 have equal \textit{pass@$o$} and \textit{pass@$k$} scores, while the other 12 LLMs (e.g., Phi3-medium-4k-instruct) show lower \textit{pass@$o$} scores. GPT-4o mini, in particular, exhibits the largest gaps on \textit{pass@$1$}, \textit{pass@$8$}, and \textit{pass@$15$} (16.21, 19.22, and 19.84). Similar results are observed across C++, C\#, PHP, and JavaScript: most LLMs, including Llama3-8b, perform worse on \textit{pass@$o$}. These findings indicate that while LLMs can sometimes generate executable code, their grasp of core OOP concepts remains weak, underscoring their limitations in capturing and applying the OOP paradigm in code generation.

\noindent\textbf{Answer to RQ3}: 
Most LLMs score lower on the \textit{pass@$o$} metric than on the \textit{pass@$k$} metric in the MultiOOP benchmark. This indicates that while these LLMs can generate executable code snippets, they do not fully understand the concepts and features of OOP.


\subsection{RQ4: High Frequency vs. Low Frequency}
Based on the classification of programming language usage frequencies in section~\ref{sec:PL_selection} and the evaluation results in Table~\ref{tab:result_01}, we can clearly see that LLMs perform better on some low-frequency programming languages than on high-frequency ones. 
For example, Qwen2.5-14b-Instruct achieves higher scores on PHP (\textit{pass@$1$}: $2.35$; \textit{pass@$8$}: $6.19$; \textit{pass@$15$}: $7.49$) than on Java (\textit{pass@$1$}: $2.05$; \textit{pass@$8$}: $6.01$; \textit{pass@$15$}: $7.12$). Similar trends are observed in other mainstream LLMs such as Llama3-8b, GPT-4o mini, CodeLlama-7b, and CodeLlama-13b.
This finding indicates that the performance of LLMs on high-frequency programming languages is sometimes not superior to that on low-frequency programming languages.

\noindent\textbf{Answer to RQ4}: The performance of LLMs in OOP tasks does not show a positive correlation with the frequency of programming language usage. Specifically, LLMs sometimes perform no better on high-frequency programming languages than on low-frequency ones. This indirectly shows that the proficiency of LLMs in a programming language may depend more on the diversity and quality of their training data than on the real-world usage frequency of the target language.

\subsection{RQ5: Prompting Selection}
In this section, we select two representative general LLMs and one code LLM as experimental subjects. Using zero-shot CoT prompting and few-shot prompting methods, we conducte systematic experiments on MultiOOP benchmark. The experimental results are shown in Table~\ref{tab:prompt_selection}. According to the experimental data in Table~\ref{tab:prompt_selection}, the few-shot prompting method demonstrates a significant performance improvement over the zero-shot prompting method in the MultiOOP benchmark. Specifically, in the Python programming environment, the Qwen2.5-14b-Instruct, StarCoder2, and GPT-4o mini models achieved \textit{pass@$1$} scores of 50.09, 33.71, and 58.40, respectively, when using the few-shot prompting method. Compared to their performance under the zero-shot prompting method, these scores represent performance improvements of 94.37\%, 1.12\%, and 21.51\%, respectively. This improvement trend is also validated by experimental results in other programming languages, such as C++ and Java.
In addition, the zero-shot CoT prompting method has demonstrated significant performance improvements in certain programming languages. Taking PHP as an example, the GPT-4o mini model, when using zero-shot CoT prompting, achieved a \textit{pass@$1$} score of 11.58, marking a 1.74\% improvement compared to the zero-shot prompting. These findings further confirm that commonly used prompting methods can also yield substantial enhancements in MultiOOP benchmark.

\noindent\textbf{Answer to RQ5}: 
LLMs with few-shot prompting show significant performance gains on the MultiOOP benchmark, while those using zero-shot CoT prompting achieve notable improvements only in specific programming languages.

\subsection{RQ6: Bad Case Analysis}
We analyze and compile the error logs generated by four high-performing LLMs (i.e., Qwen2.5-14b-Instruct, Phi3-medium-4k-instruct, StarCoder2, and GPT-4o mini) for programming languages such as Python and C++. We then present the overall error distribution of these models in Figure~\ref{fig:error_distribution}. Error types and examples in other programming languages are provided in Appendix H of the supplementary material.
From Figure~\ref{fig:error_distribution}, it is clear that common errors in Python for these LLMs include \textit{AssertionError}, \textit{AttributeError}, and \textit{TypeError}, while in C++, the primary errors are \textit{Function Call Mismatch}, \textit{Type Error}, and \textit{Undeclared Identifier}. This finding indicates that 1) LLMs exhibit varying levels of proficiency across different programming languages, and 2) there are still limitations in their understanding and generation of programming languages.

\noindent\textbf{Answer to RQ6}: 
Common errors in Python for LLMs include \textit{AssertionError}, \textit{AttributeError}, and \textit{TypeError}, while in C++, common errors include \textit{Function Call Mismatch}, \textit{Type Error}, and \textit{Undeclared Identifier}.

\subsection{RQ7: The Impact of Test Cases}
To investigate the impact of the number of test cases on the performance evaluation of LLMs in the MultiOOP task, we select GPT-4o mini and Phi3-medium-4k-instruct as reference LLMs for experimental analysis. By systematically increasing the number of test cases, we evaluate the performance trends of GPT-4o mini and Phi3-medium-4k-instruct in the MultiOOP benchmark. The experimental results, shown in Figure~\ref{tab:GPT-4o mini_performance_trend}, indicate that as the number of test cases increases, the performance evaluation scores of GPT-4o mini and Phi3-medium-4k-instruct exhibit a significant downward trend. However, when the number of test cases reaches $15$, the decline in scores tends to level off.
This trend indicates that increasing the number of test cases reduces the likelihood of accidental passes, which in turn leads to a more accurate assessment of the true performance of LLMs in the MultiOOP benchmark and yields more reliable evaluation results.

\noindent\textbf{Answer to RQ7}: Using a larger number of test cases can effectively reduce the likelihood of accidental passes, thus more accurately reflecting the true performance of the LLM.

\section{THREATS TO VALIDITY}
In the process of constructing a MultiOOP benchmark, we face two main threats. First, \textbf{the potential risk of data contamination between MultiOOP data and the pre-training data of LLMs poses a significant threat}. To address this, we design a programming language translator aimed at efficiently and accurately transforming existing OOP benchmark data into the MultiOOP variant. This method ensures data diversity while effectively mitigating the risk of data contamination. Second, \textbf{the inherent differences between programming languages present a threat to the development of a MultiOOP benchmark}. To tackle this issue, we conduct a systematic screening and filtering of existing OOP benchmark, eliminating samples that could introduce bias or misunderstanding due to linguistic differences. Following a rigorous selection process, we ultimately retain 267 representative and broadly applicable samples (see section~\ref{sec:filtering}), laying a solid data foundation for research on the MultiOOP benchmark.

\section{CONCLUSION}
In this work, we carefully design a programming language translator to convert single-language OOP benchmark into multiple programming languages. The translated OOP benchmark (namely MultiOOP) include Python, PHP, C++, C\#, Java, and JavaScript, with each language containing 267 samples.
Moreover, we enhance the \textit{pass@$o$} metric to improve its applicability across languages.
Additionally, we introduce an automated test case generation framework that systematically produces diverse test cases, enhancing the accuracy and reliability of the evaluation.
Our evaluation of 14 LLMs reveals that 1) their performance on MultiOOP tasks is notably weaker than on function-level or statement-level programming tasks, with no clear correlation to language popularity; 2) few-shot prompting substantially boosts LLM's performance, while increasing test cases effectively reduces incidental passing, ensuring a more rigorous assessment. MultiOOP benchmark fills a critical gap in multilingual OOP evaluation, offering a fair and comprehensive tool to advance research in LLM-driven code generation. In future work, we aim to optimize the translation pipeline and extend the benchmark to support more programming languages and more complex scenarios, thereby advancing the application of LLMs in software development.

\bibliography{trans}

\begin{thebibliography}{10}
\providecommand{\url}[1]{#1}
\csname url@samestyle\endcsname
\providecommand{\newblock}{\relax}
\providecommand{\bibinfo}[2]{#2}
\providecommand{\BIBentrySTDinterwordspacing}{\spaceskip=0pt\relax}
\providecommand{\BIBentryALTinterwordstretchfactor}{4}
\providecommand{\BIBentryALTinterwordspacing}{\spaceskip=\fontdimen2\font plus
\BIBentryALTinterwordstretchfactor\fontdimen3\font minus \fontdimen4\font\relax}
\providecommand{\BIBforeignlanguage}[2]{{%
\expandafter\ifx\csname l@#1\endcsname\relax
\typeout{** WARNING: IEEEtran.bst: No hyphenation pattern has been}%
\typeout{** loaded for the language `#1'. Using the pattern for}%
\typeout{** the default language instead.}%
\else
\language=\csname l@#1\endcsname
\fi
#2}}
\providecommand{\BIBdecl}{\relax}
\BIBdecl

\bibitem{herbsleb2003empirical}
J.~D. Herbsleb and A.~Mockus, ``An empirical study of speed and communication in globally distributed software development,'' \emph{IEEE Transactions on software engineering}, vol.~29, no.~6, pp. 481--494, 2003.

\bibitem{mockus2002two}
A.~Mockus, R.~T. Fielding, and J.~D. Herbsleb, ``Two case studies of open source software development: Apache and mozilla,'' \emph{ACM Transactions on Software Engineering and Methodology (TOSEM)}, vol.~11, no.~3, pp. 309--346, 2002.

\bibitem{qian2024chatdev}
C.~Qian, W.~Liu, H.~Liu, N.~Chen, Y.~Dang, J.~Li, C.~Yang, W.~Chen, Y.~Su, X.~Cong \emph{et~al.}, ``Chatdev: Communicative agents for software development,'' in \emph{Proceedings of the 62nd Annual Meeting of the Association for Computational Linguistics (Volume 1: Long Papers)}, 2024, pp. 15\,174--15\,186.

\bibitem{yang2024multi}
H.~Yang, Y.~Nong, S.~Wang, and H.~Cai, ``Multi-language software development: Issues, challenges, and solutions,'' \emph{IEEE Transactions on Software Engineering}, 2024.

\bibitem{iyer2018mapping}
S.~Iyer, I.~Konstas, A.~Cheung, and L.~Zettlemoyer, ``Mapping language to code in programmatic context,'' in \emph{Proceedings of the 2018 Conference on Empirical Methods in Natural Language Processing}, 2018, pp. 1643--1652.

\bibitem{yin2018learning}
P.~Yin, B.~Deng, E.~Chen, B.~Vasilescu, and G.~Neubig, ``Learning to mine aligned code and natural language pairs from stack overflow,'' in \emph{Proceedings of the 15th international conference on mining software repositories}, 2018, pp. 476--486.

\bibitem{chen2021evaluating}
M.~Chen, J.~Tworek, H.~Jun, Q.~Yuan, H.~P. D.~O. Pinto, J.~Kaplan, H.~Edwards, Y.~Burda, N.~Joseph, G.~Brockman \emph{et~al.}, ``Evaluating large language models trained on code,'' \emph{arXiv preprint arXiv:2107.03374}, 2021.

\bibitem{austin2021program}
J.~Austin, A.~Odena, M.~Nye, M.~Bosma, H.~Michalewski, D.~Dohan, E.~Jiang, C.~Cai, M.~Terry, Q.~Le \emph{et~al.}, ``Program synthesis with large language models,'' \emph{arXiv preprint arXiv:2108.07732}, 2021.

\bibitem{hendrycks2021measuring}
D.~Hendrycks, S.~Basart, S.~Kadavath, M.~Mazeika, A.~Arora, E.~Guo, C.~Burns, S.~Puranik, H.~He, D.~Song \emph{et~al.}, ``Measuring coding challenge competence with apps,'' \emph{arXiv preprint arXiv:2105.09938}, 2021.

\bibitem{li2022competition}
Y.~Li, D.~Choi, J.~Chung, N.~Kushman, J.~Schrittwieser, R.~Leblond, T.~Eccles, J.~Keeling, F.~Gimeno, A.~Dal~Lago \emph{et~al.}, ``Competition-level code generation with alphacode,'' \emph{Science}, vol. 378, no. 6624, pp. 1092--1097, 2022.

\bibitem{chandel2022training}
S.~Chandel, C.~B. Clement, G.~Serrato, and N.~Sundaresan, ``Training and evaluating a jupyter notebook data science assistant,'' \emph{arXiv preprint arXiv:2201.12901}, 2022.

\bibitem{athiwaratkun2023multi}
B.~Athiwaratkun, S.~K. Gouda, Z.~Wang, X.~Li, Y.~Tian, M.~Tan, W.~U. Ahmad, S.~Wang, Q.~Sun, M.~Shang \emph{et~al.}, ``Multi-lingual evaluation of code generation models,'' in \emph{The Eleventh International Conference on Learning Representations}, 2023.

\bibitem{cassano2023multipl}
F.~Cassano, J.~Gouwar, D.~Nguyen, S.~Nguyen, L.~Phipps-Costin, D.~Pinckney, M.-H. Yee, Y.~Zi, C.~J. Anderson, M.~Q. Feldman \emph{et~al.}, ``Multipl-e: a scalable and polyglot approach to benchmarking neural code generation,'' \emph{IEEE Transactions on Software Engineering}, vol.~49, no.~7, pp. 3675--3691, 2023.

\bibitem{zan2022cert}
D.~Zan, B.~Chen, D.~Yang, Z.~Lin, M.~Kim, B.~Guan, Y.~Wang, W.~Chen, and J.-G. Lou, ``Cert: continual pre-training on sketches for library-oriented code generation,'' \emph{arXiv preprint arXiv:2206.06888}, 2022.

\bibitem{zan2022language}
D.~Zan, B.~Chen, Z.~Lin, B.~Guan, W.~Yongji, and J.-G. Lou, ``When language model meets private library,'' in \emph{Findings of the Association for Computational Linguistics: EMNLP 2022}, 2022, pp. 277--288.

\bibitem{zheng2023codegeex}
Q.~Zheng, X.~Xia, X.~Zou, Y.~Dong, S.~Wang, Y.~Xue, L.~Shen, Z.~Wang, A.~Wang, Y.~Li \emph{et~al.}, ``Codegeex: A pre-trained model for code generation with multilingual benchmarking on humaneval-x,'' in \emph{Proceedings of the 29th ACM SIGKDD Conference on Knowledge Discovery and Data Mining}, 2023, pp. 5673--5684.

\bibitem{liu2024your}
J.~Liu, C.~S. Xia, Y.~Wang, and L.~Zhang, ``Is your code generated by chatgpt really correct? rigorous evaluation of large language models for code generation,'' \emph{Advances in Neural Information Processing Systems}, vol.~36, 2024.

\bibitem{liu2024multi}
Z.~Liu, G.~Zheng, and Y.~Yu, ``Multi-scale traffic pattern bank for cross-city few-shot traffic forecasting,'' \emph{arXiv preprint arXiv:2402.00397}, 2024.

\bibitem{srivastava2022beyond}
A.~Srivastava, A.~Rastogi, A.~Rao, A.~A.~M. Shoeb, A.~Abid, A.~Fisch, A.~R. Brown, A.~Santoro, A.~Gupta, A.~Garriga-Alonso \emph{et~al.}, ``Beyond the imitation game: Quantifying and extrapolating the capabilities of language models,'' \emph{arXiv preprint arXiv:2206.04615}, 2022.

\bibitem{wang2023execution}
Z.~Wang, S.~Zhou, D.~Fried, and G.~Neubig, ``Execution-based evaluation for open-domain code generation,'' in \emph{Findings of the Association for Computational Linguistics: EMNLP 2023}, 2023, pp. 1271--1290.

\bibitem{fu2023codeapex}
L.~Fu, H.~Chai, S.~Luo, K.~Du, W.~Zhang, L.~Fan, J.~Lei, R.~Rui, J.~Lin, Y.~Fang \emph{et~al.}, ``Codeapex: A bilingual programming evaluation benchmark for large language models,'' \emph{arXiv preprint arXiv:2309.01940}, 2023.

\bibitem{lai2023ds}
Y.~Lai, C.~Li, Y.~Wang, T.~Zhang, R.~Zhong, L.~Zettlemoyer, W.-t. Yih, D.~Fried, S.~Wang, and T.~Yu, ``Ds-1000: A natural and reliable benchmark for data science code generation,'' in \emph{International Conference on Machine Learning}.\hskip 1em plus 0.5em minus 0.4em\relax PMLR, 2023, pp. 18\,319--18\,345.

\bibitem{yu2024codereval}
H.~Yu, B.~Shen, D.~Ran, J.~Zhang, Q.~Zhang, Y.~Ma, G.~Liang, Y.~Li, Q.~Wang, and T.~Xie, ``Codereval: A benchmark of pragmatic code generation with generative pre-trained models,'' in \emph{Proceedings of the 46th IEEE/ACM International Conference on Software Engineering}, 2024, pp. 1--12.

\bibitem{li2023taco}
R.~Li, J.~Fu, B.-W. Zhang, T.~Huang, Z.~Sun, C.~Lyu, G.~Liu, Z.~Jin, and G.~Li, ``Taco: Topics in algorithmic code generation dataset,'' \emph{arXiv preprint arXiv:2312.14852}, 2023.

\bibitem{wang-etal-2024-oop}
S.~Wang, L.~Ding, L.~Shen, Y.~Luo, B.~Du, and D.~Tao, ``{OOP}: Object-oriented programming evaluation benchmark for large language models,'' in \emph{Findings of the Association for Computational Linguistics: ACL 2024}, 2024, pp. 13\,619--13\,639.

\bibitem{cao2024javabench}
J.~Cao, Z.~Chen, J.~Wu, S.-C. Cheung, and C.~Xu, ``Javabench: A benchmark of object-oriented code generation for evaluating large language models,'' in \emph{Proceedings of the 39th IEEE/ACM International Conference on Automated Software Engineering}, 2024, pp. 870--882.

\bibitem{lievocodebench}
J.~Li, G.~Li, X.~Zhang, Y.~Zhao, Y.~Dong, Z.~Jin, B.~Li, F.~Huang, and Y.~Li, ``Evocodebench: An evolving code generation benchmark with domain-specific evaluations,'' in \emph{The Thirty-eight Conference on Neural Information Processing Systems Datasets and Benchmarks Track}.

\bibitem{haller2024pecc}
P.~Haller, J.~Golde, and A.~Akbik, ``Pecc: Problem extraction and coding challenges,'' in \emph{Proceedings of the 2024 Joint International Conference on Computational Linguistics, Language Resources and Evaluation (LREC-COLING 2024)}, 2024, pp. 12\,690--12\,699.

\bibitem{xie2024codebenchgen}
Y.~Xie, A.~Xie, D.~Sheth, P.~Liu, D.~Fried, and C.~Rose, ``Codebenchgen: Creating scalable execution-based code generation benchmarks,'' \emph{arXiv preprint arXiv:2404.00566}, 2024.

\bibitem{babe2023studenteval}
H.~M. Babe, S.~Nguyen, Y.~Zi, A.~Guha, M.~Q. Feldman, and C.~J. Anderson, ``Studenteval: a benchmark of student-written prompts for large language models of code,'' \emph{arXiv preprint arXiv:2306.04556}, 2023.

\bibitem{zhang2024benchmarking}
Y.~Zhang, Q.~Jiang, X.~Han, N.~Chen, Y.~Yang, and K.~Ren, ``Benchmarking data science agents,'' \emph{arXiv preprint arXiv:2402.17168}, 2024.

\bibitem{li2024evocodebench}
J.~Li, G.~Li, X.~Zhang, Y.~Zhao, Y.~Dong, Z.~Jin, B.~Li, F.~Huang, and Y.~Li, ``Evocodebench: An evolving code generation benchmark with domain-specific evaluations,'' \emph{arXiv preprint arXiv:2410.22821}, 2024.

\bibitem{huang2024code}
Y.~Huang, J.~Luo, Y.~Yu, Y.~Zhang, F.~Lei, Y.~Wei, S.~He, L.~Huang, X.~Liu, J.~Zhao \emph{et~al.}, ``Da-code: Agent data science code generation benchmark for large language models,'' in \emph{Proceedings of the 2024 Conference on Empirical Methods in Natural Language Processing}, 2024, pp. 13\,487--13\,521.

\bibitem{zhuo2024bigcodebench}
T.~Y. Zhuo, M.~C. Vu, J.~Chim, H.~Hu, W.~Yu, R.~Widyasari, I.~N.~B. Yusuf, H.~Zhan, J.~He, I.~Paul \emph{et~al.}, ``Bigcodebench: Benchmarking code generation with diverse function calls and complex instructions,'' \emph{CoRR}, 2024.

\bibitem{jain2024livecodebench}
N.~Jain, K.~Han, A.~Gu, W.-D. Li, F.~Yan, T.~Zhang, S.~Wang, A.~Solar-Lezama, K.~Sen, and I.~Stoica, ``Livecodebench: Holistic and contamination free evaluation of large language models for code,'' \emph{arXiv preprint arXiv:2403.07974}, 2024.

\bibitem{xia2025leetcodedataset}
Y.~Xia, W.~Shen, Y.~Wang, J.~K. Liu, H.~Sun, S.~Wu, J.~Hu, and X.~Xu, ``Leetcodedataset: A temporal dataset for robust evaluation and efficient training of code llms,'' \emph{arXiv preprint arXiv:2504.14655}, 2025.

\bibitem{hirschberg2015advances}
J.~Hirschberg and C.~D. Manning, ``Advances in natural language processing,'' \emph{Science}, vol. 349, no. 6245, pp. 261--266, 2015.

\bibitem{otter2020survey}
D.~W. Otter, J.~R. Medina, and J.~K. Kalita, ``A survey of the usages of deep learning for natural language processing,'' \emph{IEEE transactions on neural networks and learning systems}, vol.~32, no.~2, pp. 604--624, 2020.

\bibitem{zhong2022toward}
Q.~Zhong, L.~Ding, Y.~Zhan, Y.~Qiao, Y.~Wen, L.~Shen, J.~Liu, B.~Yu, B.~Du, Y.~Chen \emph{et~al.}, ``Toward efficient language model pretraining and downstream adaptation via self-evolution: A case study on superglue,'' \emph{arXiv preprint arXiv:2212.01853}, 2022.

\bibitem{zan2022vega}
C.~Zan, K.~Peng, L.~Ding, B.~Qiu, B.~Liu, S.~He, Q.~Lu, Z.~Zhang, C.~Liu, W.~Liu \emph{et~al.}, ``Vega-mt: The jd explore academy machine translation system for wmt22,'' in \emph{Proceedings of the Seventh Conference on Machine Translation (WMT)}, 2022, pp. 411--422.

\bibitem{treviso2023efficient}
M.~Treviso, J.-U. Lee, T.~Ji, B.~v. Aken, Q.~Cao, M.~R. Ciosici, M.~Hassid, K.~Heafield, S.~Hooker, C.~Raffel \emph{et~al.}, ``Efficient methods for natural language processing: A survey,'' \emph{Transactions of the Association for Computational Linguistics}, vol.~11, pp. 826--860, 2023.

\bibitem{zhong2023can}
Q.~Zhong, L.~Ding, J.~Liu, B.~Du, and D.~Tao, ``Can chatgpt understand too? a comparative study on chatgpt and fine-tuned bert,'' \emph{arXiv preprint arXiv:2302.10198}, 2023.

\bibitem{peng2023towards}
K.~Peng, L.~Ding, Q.~Zhong, L.~Shen, X.~Liu, M.~Zhang, Y.~Ouyang, and D.~Tao, ``Towards making the most of chatgpt for machine translation,'' in \emph{Findings of the Association for Computational Linguistics: EMNLP 2023}, 2023, pp. 5622--5633.

\bibitem{achiam2023gpt}
J.~Achiam, S.~Adler, S.~Agarwal, L.~Ahmad, I.~Akkaya, F.~L. Aleman, D.~Almeida, J.~Altenschmidt, S.~Altman, S.~Anadkat \emph{et~al.}, ``Gpt-4 technical report,'' \emph{arXiv preprint arXiv:2303.08774}, 2023.

\bibitem{touvron2023llama}
H.~Touvron, L.~Martin, K.~Stone, P.~Albert, A.~Almahairi, Y.~Babaei, N.~Bashlykov, S.~Batra, P.~Bhargava, S.~Bhosale \emph{et~al.}, ``Llama 2: Open foundation and fine-tuned chat models,'' \emph{arXiv preprint arXiv:2307.09288}, 2023.

\bibitem{dubey2024llama}
A.~Dubey, A.~Jauhri, A.~Pandey, A.~Kadian, A.~Al-Dahle, A.~Letman, A.~Mathur, A.~Schelten, A.~Yang, A.~Fan \emph{et~al.}, ``The llama 3 herd of models,'' \emph{arXiv preprint arXiv:2407.21783}, 2024.

\bibitem{briot1998concurrency}
J.-P. Briot, R.~Guerraoui, and K.-P. Lohr, ``Concurrency and distribution in object-oriented programming,'' \emph{ACM Computing Surveys (CSUR)}, vol.~30, no.~3, pp. 291--329, 1998.

\bibitem{moon1986object}
D.~A. Moon, ``Object-oriented programming with flavors,'' in \emph{Conference proceedings on Object-oriented programming systems, languages and applications}, 1986, pp. 1--8.

\bibitem{madsen1989virtual}
O.~L. Madsen and B.~Moller-Pedersen, ``Virtual classes: A powerful mechanism in object-oriented programming,'' in \emph{Conference proceedings on Object-oriented programming systems, languages and applications}, 1989, pp. 397--406.

\bibitem{wilde1992maintenance}
N.~Wilde and R.~Huitt, ``Maintenance support for object-oriented programs,'' \emph{IEEE Transactions on Software Engineering}, vol.~18, no.~12, p. 1038, 1992.

\bibitem{dong2024self}
Y.~Dong, X.~Jiang, Z.~Jin, and G.~Li, ``Self-collaboration code generation via chatgpt,'' \emph{ACM Transactions on Software Engineering and Methodology}, vol.~33, no.~7, pp. 1--38, 2024.

\bibitem{wang2023review}
J.~Wang and Y.~Chen, ``A review on code generation with llms: Application and evaluation,'' in \emph{2023 IEEE International Conference on Medical Artificial Intelligence (MedAI)}.\hskip 1em plus 0.5em minus 0.4em\relax IEEE, 2023, pp. 284--289.

\bibitem{wang2024mathbb}
S.~Wang, L.~Ding, L.~Shen, Y.~Luo, Z.~He, W.~Yu, and D.~Tao, ``Improving code generation of llms by uncertainty-aware selective contrastive decoding,'' \emph{arXiv preprint arXiv:2409.05923}, 2024.

\bibitem{team2023gemini}
G.~Team, R.~Anil, S.~Borgeaud, J.-B. Alayrac, J.~Yu, R.~Soricut, J.~Schalkwyk, A.~M. Dai, A.~Hauth, K.~Millican \emph{et~al.}, ``Gemini: a family of highly capable multimodal models,'' \emph{arXiv preprint arXiv:2312.11805}, 2023.

\bibitem{team2024gemini}
G.~Team, P.~Georgiev, V.~I. Lei, R.~Burnell, L.~Bai, A.~Gulati, G.~Tanzer, D.~Vincent, Z.~Pan, S.~Wang \emph{et~al.}, ``Gemini 1.5: Unlocking multimodal understanding across millions of tokens of context,'' \emph{arXiv preprint arXiv:2403.05530}, 2024.

\bibitem{brown2020language}
T.~B. Brown, B.~Mann, N.~Ryder, M.~Subbiah, J.~Kaplan, P.~Dhariwal, A.~Neelakantan, P.~Shyam, G.~Sastry, A.~Askell \emph{et~al.}, ``Language models are few-shot learners,'' in \emph{Proceedings of the 34th International Conference on Neural Information Processing Systems}, 2020, pp. 1877--1901.

\bibitem{papineni2002bleu}
K.~Papineni, S.~Roukos, T.~Ward, and W.-J. Zhu, ``Bleu: a method for automatic evaluation of machine translation,'' in \emph{Proceedings of the 40th annual meeting of the Association for Computational Linguistics}, 2002, pp. 311--318.

\bibitem{lin2005recall}
C.~Lin, ``Recall-oriented understudy for gisting evaluation (rouge),'' \emph{Retrieved August}, vol.~20, p. 2005, 2005.

\bibitem{banerjee2005meteor}
S.~Banerjee and A.~Lavie, ``Meteor: An automatic metric for mt evaluation with improved correlation with human judgments,'' in \emph{Proceedings of the acl workshop on intrinsic and extrinsic evaluation measures for machine translation and/or summarization}, 2005, pp. 65--72.

\bibitem{wei2022chain}
J.~Wei, X.~Wang, D.~Schuurmans, M.~Bosma, F.~Xia, E.~Chi, Q.~V. Le, D.~Zhou \emph{et~al.}, ``Chain-of-thought prompting elicits reasoning in large language models,'' \emph{Advances in neural information processing systems}, vol.~35, pp. 24\,824--24\,837, 2022.

\bibitem{roziere2023code}
B.~Roziere, J.~Gehring, F.~Gloeckle, S.~Sootla, I.~Gat, X.~E. Tan, Y.~Adi, J.~Liu, R.~Sauvestre, T.~Remez \emph{et~al.}, ``Code llama: Open foundation models for code,'' \emph{arXiv preprint arXiv:2308.12950}, 2023.

\bibitem{denmat2005data}
T.~Denmat, M.~Ducass{\'e}, and O.~Ridoux, ``Data mining and cross-checking of execution traces: a re-interpretation of jones, harrold and stasko test information,'' in \emph{Proceedings of the 20th IEEE/ACM international Conference on Automated software engineering}, 2005, pp. 396--399.

\bibitem{grote2008hybridizing}
M.~Grote, ``Hybridizing bacteria, crossing methods, cross-checking arguments: The transition from episomes to plasmids (1961-1969),'' \emph{History and philosophy of the life sciences}, pp. 407--430, 2008.

\bibitem{abdin2024phi}
M.~Abdin, J.~Aneja, H.~Awadalla, A.~Awadallah, A.~A. Awan, N.~Bach, A.~Bahree, A.~Bakhtiari, J.~Bao, H.~Behl \emph{et~al.}, ``Phi-3 technical report: A highly capable language model locally on your phone,'' \emph{arXiv preprint arXiv:2404.14219}, 2024.

\bibitem{yang2024qwen2}
A.~Yang, B.~Yang, B.~Zhang, B.~Hui, B.~Zheng, B.~Yu, C.~Li, D.~Liu, F.~Huang, H.~Wei \emph{et~al.}, ``Qwen2. 5 technical report,'' \emph{arXiv preprint arXiv:2412.15115}, 2024.

\bibitem{li2023starcoder}
R.~Li, L.~B. Allal, Y.~Zi, N.~Muennighoff, D.~Kocetkov, C.~Mou, M.~Marone, C.~Akiki, J.~Li, J.~Chim \emph{et~al.}, ``Starcoder: may the source be with you!'' \emph{arXiv preprint arXiv:2305.06161}, 2023.

\bibitem{lozhkov2024starcoder}
A.~Lozhkov, R.~Li, L.~B. Allal, F.~Cassano, J.~Lamy-Poirier, N.~Tazi, A.~Tang, D.~Pykhtar, J.~Liu, Y.~Wei \emph{et~al.}, ``Starcoder 2 and the stack v2: The next generation,'' \emph{arXiv preprint arXiv:2402.19173}, 2024.

\bibitem{xu2023wizardlm}
C.~Xu, Q.~Sun, K.~Zheng, X.~Geng, P.~Zhao, J.~Feng, C.~Tao, and D.~Jiang, ``Wizardlm: Empowering large language models to follow complex instructions,'' \emph{arXiv preprint arXiv:2304.12244}, 2023.

\end{thebibliography}
\bibliographystyle{IEEEtran}
\clearpage
\appendices

\twocolumn[ 
\vspace{1em} 
\begin{center}
    {\LARGE \textbf{Supplementary Material}\par} 
    \vspace{1em}
\end{center}
]

\noindent
In this file, we provide:

\noindent
1. Detailed process of programming language (C++, Java, C\#, PHP, Python, and JavaScript) classification.

\noindent
2. Detailed description of the OOP benchmark~\cite{wang-etal-2024-oop} filtering process.

\noindent
3. Detailed process of translating the single-language OOP benchmark into a multilingual setting.

\noindent
4. Prompts design for automated augmentation of test cases.

\noindent
5. Detailed description of the evaluated LLMs.

\noindent
6. Detailed description of the \textit{pass@$k$} and \textit{pass@$o$} metrics.

\noindent
7. Further supplementation of the experimental results.

\noindent
8. Further supplementation of the bad case analysis.

\noindent
9. Design considerations regarding the number of test cases in the MultiOOP benchmark.

\section{The classification of programming language frequency}
Based on programming language rankings from four major platforms (see Table II), we reassess the selected languages. For example, on the GitHut 2.0 platform, the usage shares of C++, Java, C\#, PHP, Python, and JavaScript are 9.49\%, 11.75\%, 3.45\%, 5.69\%, 16.99\%, and 9.90\%, respectively. Accordingly, we rank them on the GitHut 2.0 platform as follows: Python (1st), Java (2nd), JavaScript (3rd), C++ (4th), PHP (5th), and C\# (6th). We perform similar re-rankings on the TIOBE, PYPL, and RedMonk platforms, with the detailed results shown in Table~\ref{tab:programm_language_rank}. Based on the average rankings across all platforms, we categorize the six languages into high-frequency, medium-frequency, and low-frequency programming language groups.

\setcounter{table}{5}
\begin{table}[!b]
  \centering
  \caption{The re-ranking results of the six selected programming languages (i.e., C++, Java, C\#, PHP, Python, and JavaScript).}
  \resizebox{0.98\linewidth}{!}{
    \begin{tabular}{cccccc}
    \toprule
    PL    & \multicolumn{1}{c}{GitHut 2.0} & \multicolumn{1}{c}{TIOBE} & \multicolumn{1}{c}{PYPL} & \multicolumn{1}{c}{RedMonk} & \multicolumn{1}{c}{Avg} \\
    \midrule
    C++   & 4 & 2     & 4 & 6     & 4.00 \\
    Java  & 2 & 3     & 2 & 3     & 2.50 \\
    C\#   & 6 & 4     & 5 & 5     & 5.00 \\
    PHP   & 5 & 6    & 6 & 4     & 5.25 \\
    Python & 1 & 1     & 1 & 2     & 1.25 \\
    JavaScript & 3 & 5     & 3 & 1     & 3.00 \\
    \bottomrule
    \end{tabular}%
    }
  \label{tab:programm_language_rank}%
\end{table}%

\section{Detailed Filtering Process of the OOP Benchmark}
To develop the MultiOOP benchmark, we aim to extend an existing OOP benchmark~\cite{wang-etal-2024-oop}, originally designed for a single-language (i.e., Python), to five additional programming languages (i.e., C++, Java, C\#, PHP, and JavaScript). However, due to the unique features and limitations of each language, not all code samples are suitable for cross-language translation. For instance, Python includes distinctive constructs (e.g., dynamically typed list comprehensions) that lack straightforward equivalents in statically typed languages like C++ or Java, potentially resulting in inaccurate translations. To ensure consistent and high-quality data, we rigorously select samples according to two main criteria: semantic equivalence and language-agnostic design.


\noindent
\textbf{Semantic equivalence:}
We select samples that demonstrate core object-oriented programming concepts (e.g., classes and inheritance) and are suitable for accurate translation into five target languages without any loss of functionality. Samples involving Python-style lists are excluded because such constructs lack straightforward equivalents in statically typed languages like C++ or Java, often requiring complex workarounds that result in inconsistent translations, as shown in Listing~\ref{lst:list_example}.

\begin{lstlisting}[language=Python, escapeinside={(*@}{@*)}, caption= {An example sample from the OOP benchmark that are specified to return outputs in list format.}, label={lst:list_example}]
(*@\textbf{Input:}@*)
First, write a class called (*@\textbf{AAGM}@*) using the 
Python language. Then, within the (*@\textbf{AAGM}@*) class,
create a public function called (*@\textbf{anagram}@*) that takes 
an array of strings as input. This function should 
group together anagrams and return the result as a 
list.

(*@\textbf{Python code generated by GPT-4o:}@*)
class AAGM:
    def anagram(self, strs):
        anagram_map = {}
        for s in strs:
            key = ''.join(sorted(s))
            if key not in anagram_map:
                anagram_map[key] = []
            anagram_map[key].append(s)
        # Return the grouped anagrams as a list of lists
        return list(anagram_map.values())
\end{lstlisting}

\setcounter{figure}{6}

\begin{figure}[!t]
    \centering
    \includegraphics[width=0.48\textwidth]{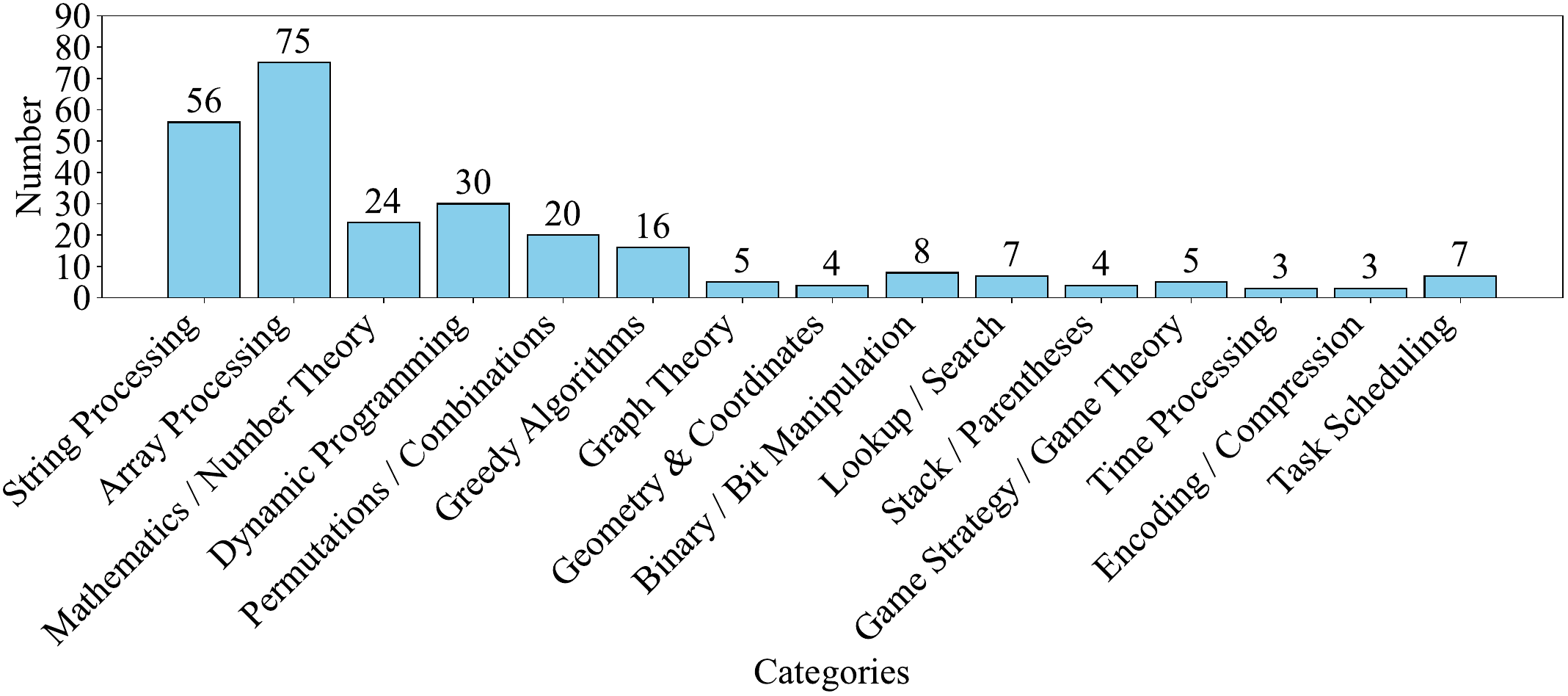}
    \caption{Distribution of categories in the filtered OOP benchmark.}
    \label{fig:category_distribution}
\end{figure}

Instead, we prioritize tasks based on universal OOP structures (e.g., class hierarchies), as they tend to translate more reliably and consistently across different languages.

\noindent
\textbf{Language-agnostic design:}
To ensure broad applicability, samples that rely on language-specific libraries or features are excluded. For example, samples that depend on Python’s ``collections''. ``Counter'' module are removed, since equivalent functionality in C++ or PHP often requires custom implementations, thereby reducing comparability.

Based on the two key criteria mentioned above, we ultimately select 267 samples. All these samples pass compatibility tests on Python, C++, Java, C\#, PHP, and JavaScript, and are manually reviewed by two programming experts. The distribution of the 267 sample types is illustrated in Figure~\ref{fig:category_distribution}.

\section{Translation Converter Implementation Details}
To extend the OOP benchmark from a single-language setting to a multilingual environment, we developed a programming language translator. This translator converts Python-based OOP benchmark code into five other programming languages, i.e., C++, Java, C\#, PHP, and JavaScript. It handles syntactic differences, ensures semantic consistency, and incorporates validation mechanisms to maintain correctness, thereby enabling filtered OOP benchmark to operate across multiple languages.

\noindent 
\textbf{Syntactic differences:}
The translator adopts a modular pipeline architecture to handle syntactic differences between various programming languages. The process begins by parsing the source code into an abstract syntax tree (AST) using language-specific parsers, such as Python's ``ast'' module or C++'s ``Clang'' module. The resulting AST is then transformed into an intermediate representation (IR), which abstracts language-specific constructs into a unified form. For example, consider the boolean function in Python, as shown in Listing~\ref{lst:boolean_example}.

\begin{lstlisting}[language=Python, caption={An example of a boolean function in Python.}, label={lst:boolean_example}]
def validate_function(code, cases):
   for case in cases:
      if case not in code:
        return False
   return True
\end{lstlisting}

This function is converted into an intermediate representation (IR) containing general information, such as the function name. Subsequently, the IR is used with predefined templates to generate code in the target language. The corresponding C++ implementation is as shown in Listing~\ref{lst:boolean_example_C}.

\begin{lstlisting}[language=C++, caption={An example of a boolean function in C++.}, label={lst:boolean_example_C}]
#include <string>
#include <vector>

bool validate_function(const std::string& code, const std::vector<std::string>& cases){
    for (const auto& case_str : cases){
        if (code.find(case_str) == std::string::npos){
            return false;
        }
    }
    return true;
}
\end{lstlisting}

To address syntactic differences between languages, we maintain a set of rule-based mapping mechanisms for each pair of source and target languages. For example, Python's ``def'' is mapped to a return type like ``bool'' in C++. These mappings are validated according to the style guides of each language to ensure that the generated code complies with the conventions and best practices of the target language.

\noindent
\textbf{Semantic consistency:}
To ensure semantic consistency, the translator verifies the functional equivalence of the intermediate representation (IR) by comparing the input and output behavior across different languages. For example, here is a test case in Python, as shown in Listing~\ref{lst:test_case_python}.

\begin{lstlisting}[language=Python, caption={An example of a test case in Python.}, label={lst:test_case_python}]
input1 = [1, 2, 4, 8, 16]
assert SN_RSF(input1).Reorganization_satisfaction()
                                            ==False
\end{lstlisting}

It will be converted into an equivalent C++ test code, as shown in Listing~\ref{lst:test_case_C}.
\begin{lstlisting}[language=C++, caption={An example of a test case in C++.}, label={lst:test_case_C}]
std::vector<std::pair<int, int>> input1 = {1, 2, 4,
                                             8, 16};
assert(SN_RSF(input1).Reorganization_satisfaction()
                                           ==False);
\end{lstlisting}

If any inconsistencies are found in the test results, the converter will iteratively refine the translation based on the feedback from the failed tests.

\noindent
\textbf{Validation:}
To ensure the accuracy of the translation results, we conduct rigorous validation tests on the converter. We perform experiments on a carefully selected subset of 50 object-oriented programming (OOP) examples, demonstrating a 98.00\% success rate in generating syntactically correct and semantically equivalent code across six languages. Additionally, by introducing type annotations in the IR, we effectively reduce issues such as type inference errors common in weakly typed languages. To further enhance translation accuracy, we evaluate the translated target code together with the source code using GPT-4o~\footnote{\url{https://platform.openai.com/docs/models/gpt-4o.}} to verify correctness. The detailed design is shown in Figure~\ref{fig:equivalence_prompt}.

\begin{figure}[!t]
\centering
\begin{tcolorbox}[colback=yellow!10!white, colframe=gray!15, coltitle=black]
\fontsize{7}{9}\selectfont You are a code analysis expert. Your task is to evaluate whether the following two code snippets are logically and functionally equivalent. If the two snippets produce the same output for any given input (i.e., they are behaviorally equivalent), return 1; otherwise, return 0.\\
\\
Please return only the number 1 or 0, without any additional explanation.\\
\\
Python code: \\
$\{$source code$\}$ \\
\\
C++ code: \\
$\{$translated target code$\}$
\end{tcolorbox}
\vspace{-1em}
\caption{Prompt for evaluating code equivalence using GPT-4o. Here, we use C++ as a reference, and the evaluation of code in other translated languages (i.e., Java, C\#, PHP, and JavaScript) is similar.}
\label{fig:equivalence_prompt}
\end{figure}


Based on the prompt shown in Figure~\ref{fig:equivalence_prompt}, we evaluate the translation results of five programming languages, with 267 samples per language, as shown in Table~\ref{tab:equivalence_evaluation}. Additionally, we hired two professional programmers to score the results using a cross-checking method~\cite{denmat2005data,grote2008hybridizing}. A score of 1 is assigned if the original code and the translated code are equivalent; otherwise, a score of 0 is given. We then average the scores from the two programmers.

Using the implementation approach and evaluation results described above, we ensure that the MultiOOP benchmark accurately reflects the capabilities of LLMs across different programming languages, while minimizing translation errors and preserving functional integrity.

\begin{table}[!t]
  \centering
  \caption{Accuracy (\%) of source code translation to target code. The ``manual'' indicates the evaluation results of professional programmers.}
  \resizebox{0.98\linewidth}{!}{
    \begin{tabular}{cccccc}
    \toprule
    \multicolumn{1}{c}{Target language} & \multicolumn{1}{c}{C++} & \multicolumn{1}{c}{Java} & \multicolumn{1}{c}{C\#} & \multicolumn{1}{c}{PHP} & \multicolumn{1}{c}{JavaScript} \\
    \midrule
    GPT-4o & 99.63     & 100.00     & 98.13     & 100.00     & 99.63 \\
    Manual & 98.88     & 99.63     & 98.13     & 100.00     & 99.51 \\
    \bottomrule
    \end{tabular}%
    }
  \label{tab:equivalence_evaluation}%
\end{table}%

\section{Details of automatically generating new test cases}
In this process, we selected GPT-4o as the model for generating new test cases and carefully designed corresponding prompts (as shown in Figure~\ref{fig:tcb_prompt}) to guide GPT-4o in generating these new test cases.

\section{A detailed description of LLMs}
In this work, we evaluate nine mainstream general LLMs (i.e., Llama3-8b, Llama3.1-8b, Llama3.1-8b-Instruct, Llama3.2-11b, Phi3-medium-4k-instruct, Qwen2.5-7b-Instruct, Qwen2.5-14b-Instruct, Vicuna-13b-v1.5, and GPT-4o mini) and five widely used code-specialized LLMs (i.e., CodeLlama-7b, CodeLlama-13b, StarCoder, StarCoder2, and WizardCoder-15b-V1.0) using the constructed MultiOOP benchmark. Below is a detailed description to these 14 LLMs:

\noindent $\bullet$ \textbf{Llama3-8b~\cite{dubey2024llama}}:
Llama3-8b is an open-source LLM developed by Meta based on the Llama2~\cite{touvron2023llama} architecture. Compared to Llama2-7b, the main improvement in Llama3-8b is the introduction of a new tokenizer, expanding its vocabulary size to 128,256 words. This expanded vocabulary not only allows for more efficient text encoding (both input and output) but may also enhance the model's ability to handle multiple languages.

\begin{figure}[!t]
\centering
\begin{tcolorbox}[colback=yellow!10!white, colframe=gray!15, coltitle=black]
\fontsize{7}{9}\selectfont You are a skilled Python programmer. Based on the test cases provided in the Python code, generate $m$ new and unique test cases. Ensure that these new test cases cover a variety of scenarios, edge cases, and input types to improve the diversity and comprehensiveness of the testing. Avoid duplicating any existing test cases.\\
\verb|```| python \\
$\{$ground-truth code$\}$ \\
$\{$existing test cases$\}$\\
\verb|```| \\
Importantly, the format of the newly generated test cases must remain consistent with the existing test cases.
\end{tcolorbox}
\vspace{-1em}
\caption{Prompts for generating new test cases}
\label{fig:tcb_prompt}
\end{figure}

\noindent $\bullet$ \textbf{Llama3.1-8b~\cite{dubey2024llama}}: Llama3.1-8b is an important member of the Llama series. Compared to the Llama3 version, the Llama3.1 version has been improved in several aspects, aiming to enhance the quality and efficiency of text generation. Trained on vast datasets, Llama3.1-8b is capable of understanding and generating natural text in multiple languages. In contrast to other LLMs, the Llama3.1-8b emphasizes efficiency and scalability, achieving near-state-of-the-art performance with relatively low computational resources.

\noindent $\bullet$ \textbf{Llama3.1-8b-Instruct~\cite{dubey2024llama}}: Llama3.1-8b is specifically optimized for multilingual dialogue scenarios and has excelled in several industry-standard benchmarks, outperforming many existing open-source and closed-source chat models.

\noindent $\bullet$ \textbf{Llama3.2-11b~\cite{dubey2024llama}}:
Llama3.2-11b is developed based on the Llama3.1 pure text model. Llama3.2-11b has been fine-tuned through supervised learning and reinforcement learning with human feedback to better meet human needs for practicality and safety. To enhance image recognition capabilities, Llama3.2-11b introduces an independently trained visual adapter, which integrates with the pre-trained Llama3.1 language model. This adapter consists of multiple cross-attention layers, allowing the output from the image encoder to be passed to the core LLM.

\noindent $\bullet$ \textbf{Phi3-medium-4k-instruct~\cite{abdin2024phi}}:
Phi3-medium-4k-Instruct is a lightweight open-source model with 14 billion parameters. The model has undergone post-training steps, including supervised fine-tuning and direct preference optimization, to ensure it follows instructions and meets safety requirements. Phi-3-Medium-4K-Instruct demonstrates industry-leading performance across multiple tasks, including language understanding, mathematical operations, programming code handling, long-context comprehension, and logical reasoning.

\begin{table*}[!t]
  \centering
  \caption{Overview of the evaluated models.}
  \resizebox{1.0\linewidth}{!}{
    \begin{tabular}{lllcll}
    \toprule
    \textbf{Model name} & \textbf{Organization}  & \multicolumn{1}{c}{\textbf{Years}}  & \multicolumn{1}{c}{\textbf{Open-source}} & \multicolumn{1}{c}{\textbf{Size}} &
    \multicolumn{1}{c}{\textbf{Source}} \\
    \midrule
    Llama3-8b & Meta & $2024$ & \checkmark & 8B & \url{https://huggingface.co/meta-llama/Meta-Llama-3-8B} \\
    Llama3.1-8b & Meta & $2024$ & \checkmark & 8B & \url{https://huggingface.co/meta-llama/Llama-3.1-8B} \\
    Llama3.1-8b-Instruct & Meta & $2024$ & \checkmark & 8B & \url{https://huggingface.co/meta-llama/Llama-3.1-8B-Instruct} \\
    Llama3.2-11b & Meta & $2024$ & \checkmark & 11B & \url{https://huggingface.co/meta-llama/Llama-3.2-11B-Vision} \\
    Phi3-medium-4k-instruct & Microsoft & $2024$ & \checkmark & 14B & \url{https://huggingface.co/microsoft/Phi-3-medium-4k-instruct} \\
    Qwen2.5-7b-Instruct & Alibaba & $2024$ & \checkmark & 7B & \url{https://huggingface.co/Qwen/Qwen2.5-7B-Instruct} \\
    Qwen2.5-14b-Instruct & Alibaba & $2024$ &  \checkmark & 14B &  \url{https://huggingface.co/Qwen/Qwen2.5-14B-Instruct} \\
    Vicuna-13b-v1.5 & UC Berkeley & $2023$ & \checkmark  & 13B & \url{https://huggingface.co/lmsys/vicuna-13b-v1.5} \\
    GPT-4o mini & Open AI & $2024$ & \ding{55}  & - & \url{https://huggingface.co/lmsys/vicuna-13b-v1.5} \\
    CodeLlama-7b & Meta & $2023$ & \checkmark  & 7B & \url{https://huggingface.co/codellama/CodeLlama-7b-hf} \\
    CodeLlama-13b & Meta & $2023$ & \checkmark  & 13B & \url{https://huggingface.co/codellama/CodeLlama-13b-hf} \\
    StarCoder & Hugging Face & $2023$ & \checkmark  & 15B & \url{https://huggingface.co/bigcode/starcoder} \\
    StarCoder2 & Hugging Face & $2024$ & \checkmark  & 15B & \url{https://huggingface.co/bigcode/starcoder2-15b} \\
    WizardCoder-15b-V1.0 & Microsoft & $2023$ & \checkmark  & 15B & \url{https://huggingface.co/WizardLMTeam/WizardCoder-15B-V1.0} \\
    \bottomrule
    \end{tabular}%
    }
  \label{tab:models_overview}%
\end{table*}%

\noindent $\bullet$ \textbf{Qwen2.5-7b-Instruct~\cite{yang2024qwen2}}:
Qwen2.5-7b-Instruct is a newly launched model by Alibaba, pretrained on a large-scale dataset containing up to 18 trillion tokens, and capable of processing up to 128K tokens and supports multiple languages.

\noindent $\bullet$ \textbf{Qwen2.5-14b-Instruct~\cite{yang2024qwen2}}:
Qwen2.5-14b-Instruct is an open-source model with 14 billion parameters, offering exceptional reasoning and contextual understanding capabilities. Compared to traditional GPT models, Qwen2.5-14b-Instruct performs more accurately and efficiently when executing specific task instructions. Especially when given clear instructions from the user, it is able to quickly generate meaningful outputs.

\noindent $\bullet$ \textbf{Vicuna-13b-v1.5~\cite{yang2024qwen2}}: Vicuna-13b-v1.5 is a model fine-tuned from the Llama2 series~\cite{touvron2023llama} through supervised instruction tuning. The fine-tuning process leverages approximately 125K conversation instances sourced from ShareGPT~\footnote{\url{https://sharegpt.com/.}}, enabling the model to acquire enhanced instruction-following capabilities.


\noindent $\bullet$ \textbf{GPT-4o mini~\footnote{\url{https://openai.com/index/gpt-4o-mini-advancing-cost-efficient-intelligence/.}}}:
GPT-4o mini is a streamlined version of the GPT-4 model developed by OpenAI. The design goal is to make it more lightweight in terms of computational resources and processing power compared to the full version of GPT-4, while still maintaining a high level of text generation quality. GPT-4o mini is especially suitable for applications that have lower computational requirements or are more sensitive to cost.

\noindent $\bullet$ \textbf{CodeLlama-7b~\cite{roziere2023code}}:
CodeLlama-7b is a text generation model developed and open-sourced by Meta, built upon the Llama2 architecture~\cite{touvron2023llama}. It is explicitly designed to address a wide range of tasks related to code generation and code comprehension.


\noindent $\bullet$ \textbf{CodeLlama-13b~\cite{roziere2023code}}:
CodeLlama-13b is a LLM based on the LLaMA architecture, with 13 billion parameters, specifically designed for tasks related to code generation, understanding, and optimization. Developed by Meta AI, the model aims to enhance developers' efficiency in writing, debugging, and maintaining code.

\begin{figure}[!t]
    \centering
    \includegraphics[width=0.45\textwidth]{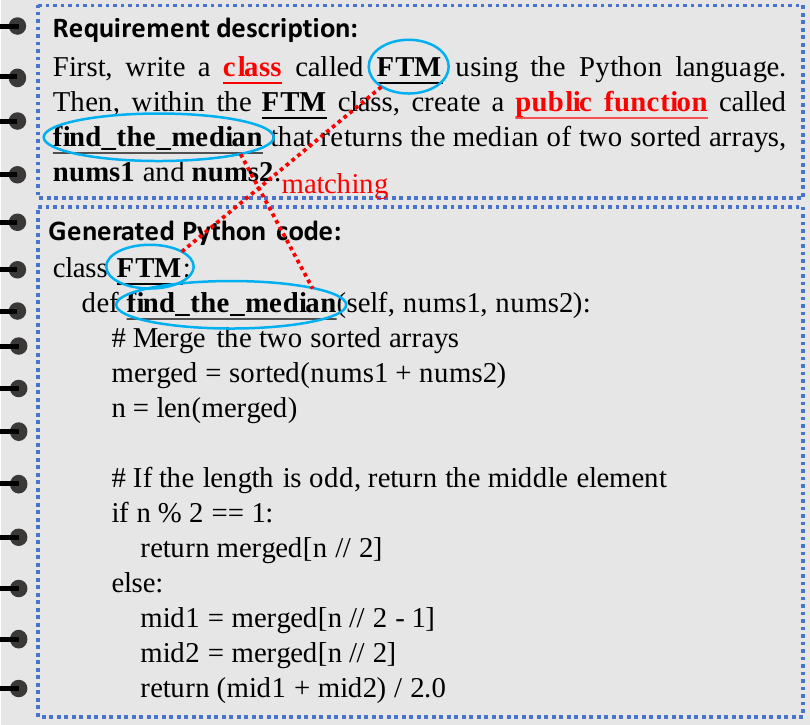}
    \caption{Illustration of matching key points between natural language description and program description in the \textit{pass@$o$} metric. The matching between the key points (i.e., \underline{\textbf{class FTM}} and \underline{\textbf{def find\_the\_median}}) in the natural language description and those in the program is shown.}
    \label{fig:matching_key_point}
\end{figure}

\begin{figure}[!t]
    \centering
    \includegraphics[width=0.45\textwidth]{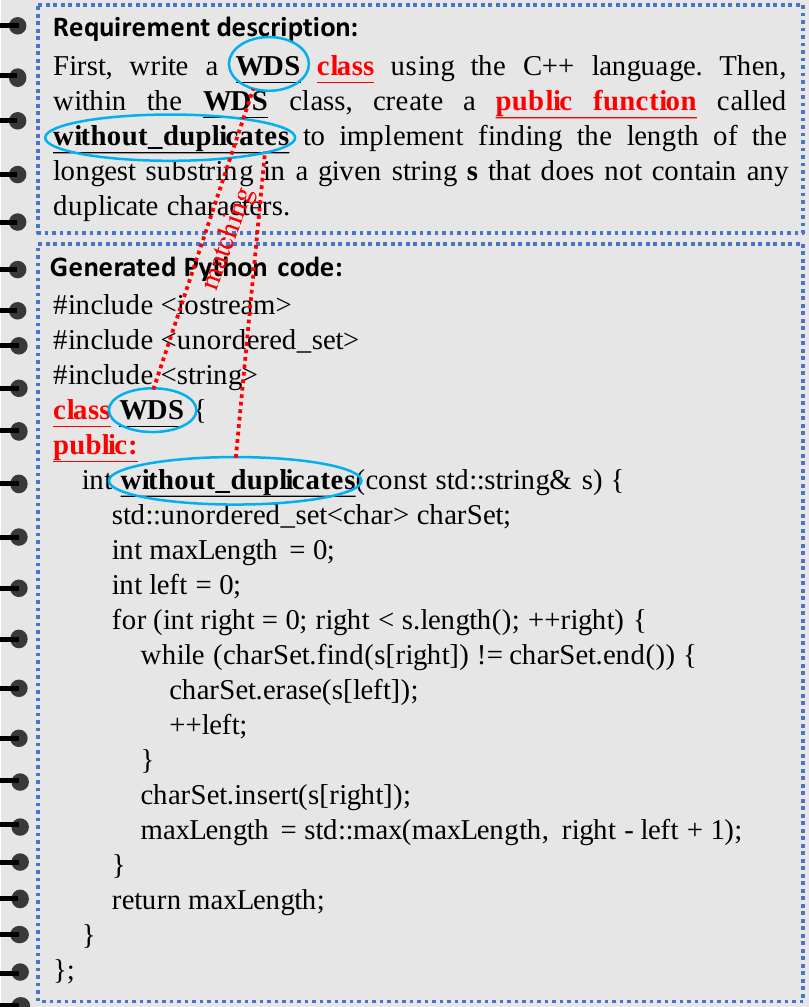}
    \caption{Illustration of matching key points between natural language description and program description in the \textit{pass@$o$} metric extended to multiple languages. Here, we take C++ as reference. The matching between the key points (i.e., \underline{\textbf{class WDS}} and \underline{\textbf{public:  int without\_duplicates}}) in the natural language description and those in the program is shown.}
    \label{fig:matching_key_point01}
\end{figure}

\noindent $\bullet$ \textbf{StarCoder~\cite{li2023starcoder}}:
StarCoder is a large-scale code language model that is trained using licensed data from GitHub~\footnote{\url{https://github.com/.}}, encompassing over 80 programming languages, Git commit records, GitHub issues, and Jupyter notebook content.

\noindent $\bullet$ \textbf{StarCoder2~\cite{lozhkov2024starcoder}}:
StarCoder2 is an open-source large language model designed specifically for coding. It is based on the Stack v2 dataset~\footnote{\url{https://huggingface.co/datasets/bigcode/the-stack-v2.}} and has been trained on over 4 trillion tokens across more than 600 programming languages.

\noindent $\bullet$ \textbf{WizardCoder-15b-V1.0~\cite{xu2023wizardlm}}: WizardCoder-15b-V1.0 is a task DAG scheduler developed based on Golang, specifically designed for handling large-scale data processing and analysis tasks. It efficiently manages and schedules tasks within the DAG, ensuring that tasks are executed in the correct order, thereby optimizing resource utilization and enhancing processing efficiency.

We have summarized the specific details of above models in Table~\ref{tab:models_overview}.

\section{A detailed description of \textit{pass@$k$} and \textit{pass@$o$}}

The \textit{pass@$k$} is a metric proposed by Chen et al.~\cite{chen2021evaluating} in 2021, used to evaluate whether code generated by LLMs can run correctly by passing a set of predefined test cases.
This \textit{pass@$k$} metric requires the LLMs to generate $n$ ($n \geq k$) code samples for each problem (i.e., requirement description), and then randomly select $k$ of these samples for testing. If at least one of the $k$ samples passes the unit test, the problem is considered to have passed the test. The calculation formula for the \textit{pass@$k$} metric is:

\begin{equation}
\label{eq:mertric_pass@k}
\textit{pass@$k$}:=\mathop{\mathbb{E}}_{Problems} \left[1-\frac{{\binom{n-c}{k}}}{\binom{n}{k}} \right]
\end{equation}
In Eq. (\ref{eq:mertric_pass@k}), \textit{c} represents the total number of all generated codes that can pass unit tests.

The \textit{pass@$o$} is a metric proposed by Wang et al.~\cite{wang-etal-2024-oop} in 2024 that builds upon the \textit{pass@$k$} metric by adding the requirement of matching key points between the natural language description and the program description, as shown in Figure~\ref{fig:matching_key_point}. This means that \textit{pass@$o$} ensures the model not only generates executable code but also correctly implements OOP concepts such as class and encapsulation, making it more suitable for evaluating the complexity of OOP tasks. The calculation formula for the \textit{pass@$o$} metric is:
\begin{align}
\label{eq:total_add_pass_matching}
&\quad\quad\quad\quad\,\,\,\,\alpha = \sum_{i=1}^{n} f\left(X_i\right),\nonumber\\
&where f(X_i)= \\
&\left\{
\begin{aligned}
    1,& \, if \, utf\left(X_i\right) \, passed \, and \, \sum_{j}^{m} x_j \exists X_i \\
    0,& \, \mathrm{otherwise}
\end{aligned}
\right., \nonumber
\end{align}

\begin{align}
\label{eq:mertric_pass@o}
\textit{pass@$o$}&:=\mathop{\mathbb{E}}_{Problems} \left[1-\frac{{\binom{n-\alpha}{k}}}{\binom{n}{k}} \right]
\end{align}
In Eq. (\ref{eq:total_add_pass_matching}), $X_i$ represents the $i$-${th}$ generated code samples; $\alpha$ represents the number of $n$ generated code samples that pass the unit test and achieve successful matching; $ut\left(\cdot\right)$ represents the unit test function; $m$ denotes the number of key points in the current natural language description; and $x_j$ denotes the $j$-${th}$ keyword points in natural language.

In this work, we extend the single-language \textit{pass@$o$} metric to multiple programming languages (as shown in Figure~\ref{fig:matching_key_point01}) to evaluate the generality and robustness of LLMs in a multilingual setting.

\begin{table*}[!t]
  \centering
  \caption{Performance of 13 mainstream large language models (LLMs) on MultiOOP tasks. We also reported the differences in evaluation results between \textit{pass@$k$} and \textit{pass@$o$}. (In the experiments, we set the temperature to 0.8, and all LLMs are evaluated in zero-shot prompting. Bold indicates the differences evaluated using the \textit{pass@$k$} and \textit{pass@$o$} metrics; Red indicates the best evaluation result; Underlined indicates the maximum disparities evaluated between \textit{pass@$k$} and \textit{pass@$o$} metrics; Gray indicates specialized code LLMs.)}
  \resizebox{1.0\linewidth}{!}{
    \begin{tabular}{l||ccccccccc||ccccccccc}
    \toprule
    \multicolumn{1}{l}{\multirow{2}[4]{*}{Model}} & \multicolumn{3}{c}{1} & \multicolumn{3}{c}{8} & \multicolumn{3}{c}{15} & \multicolumn{3}{c}{1} & \multicolumn{3}{c}{8} & \multicolumn{3}{c}{15} \\
\cmidrule{2-19}    \multicolumn{1}{c||}{} & $k$ & $o$ & $\boldsymbol{\Delta}\left(\downarrow\right)$     & $k$ & $o$ & $\boldsymbol{\Delta}\left(\downarrow\right)$     & $k$ & $o$ & \multicolumn{1}{c||}{$\boldsymbol{\Delta}\left(\downarrow\right)$} & $k$ & $o$ & $\boldsymbol{\Delta}\left(\downarrow\right)$     & $k$ & $o$ & W     & $k$ & $o$ & $\boldsymbol{\Delta}\left(\downarrow\right)$ \\
    \midrule
    \multicolumn{1}{c||}{} & \multicolumn{9}{c||}{Python (High)}                                            & \multicolumn{9}{c}{C++ (Medium)} \\
    \midrule
    Llama3-8b & 0.77  & 0.72  & \textbf{-0.05}  & 5.01  & 4.98  & \textbf{-0.03}  & 8.24  & 7.87  & \textbf{-0.37}
    & 0.07  & 0.03  & \textbf{-0.04}  & 0.80 & 0.52  & \textbf{-0.28}  & 1.47  & 1.21  & \textbf{-0.26}  \\
    Llama3.1-8b & 1.07  & 0.77  & \textbf{-0.30}  & 6.82  & 4.83  & \textbf{-1.99}  & 10.49  & 7.87  & \textbf{-2.62}
    & 1.25  & 0.75  & \textbf{-0.50}  & 6.29 & 5.35  & \textbf{-0.94}  & 9.11  & 6.99  & \textbf{-2.12}  \\
    Llama3.1-8b-Instruct & 7.53  & 6.89  & \textbf{-0.64}  & 26.43  & 21.91  & \textbf{-4.52}  & 33.18  & 26.22  & \textbf{-6.96} 
    & 1.40  & 0.92  & \textbf{-0.48}  & 6.92 & 4.81  & \textbf{-2.11}  & 9.36  & 7.12  & \textbf{-2.24}  \\
    Llama3.2-11b & 8.96  & 6.72  & \textbf{-2.24}  & 30.76  & 21.61  & \textbf{-9.15}  & 37.45  & 27.72  & \textbf{-9.73}
    & 1.37  & 1.10  & \textbf{-0.27}  & 6.13 & 3.71  & \textbf{-2.42}  & 8.24  & 4.12  & \textbf{-4.12}  \\
    Phi3-medium-4k-instruct & 21.20  & 19.58  & \textbf{-1.62}  & 48.95  & 47.09  & \textbf{-1.86}  & 55.43  & 53.56  & \textbf{-1.87} 
    & 4.29  & 3.15  & \textbf{-1.14}  & 13.64 & 10.22  & \textbf{-3.42}  & 16.85  & 12.73  & \textbf{-4.12}  \\
    Qwen2.5-7b-Instruct & 21.00  & 13.21  & \underline{\textbf{-7.79}}  & 56.66  & 38.64  & \underline{\textbf{-18.02}}  & 64.79  & 46.44  & \underline{\textbf{-18.35}}
    & 4.32  & 1.05  & \textbf{-3.27}  & 15.28 & 6.58  & \underline{\textbf{-8.70}}  & 18.35  & 9.34  & \underline{\textbf{-9.01}}  \\
    Qwen2.5-14b-Instruct & 24.59  & 18.90  & \textbf{-5.69}  & \textcolor{red}{\textbf{63.20}}  & \textcolor{red}{\textbf{54.82}}  & \textbf{-8.38}  & \textcolor{red}{\textbf{70.41}} & \textcolor{red}{\textbf{64.04}}  & \textbf{-6.37} 
    & \textcolor{red}{\textbf{10.31}}  & \textcolor{red}{\textbf{5.37}}  & \underline{\textbf{-4.94}}  & \textcolor{red}{\textbf{21.78}} & \textcolor{red}{\textbf{15.03}}  & \textbf{-6.75}  & \textcolor{red}{\textbf{25.84}}  & \textcolor{red}{\textbf{17.60}}  & \textbf{-8.24}  \\
    Vicuna-13b-v1.5 & 0.62  & 0.60  & \textbf{-0.02}  & 4.29  & 3.57  & \textbf{-0.72}  & 7.12  & 5.62  & \textbf{-1.50}
    & 0.02  & 0.02  & \textbf{-0.00} & 0.28 & 0.23  & \textbf{-0.05}  & 0.45  & 0.40  & \textbf{-0.05}  \\
    \rowcolor[rgb]{ .906,  .902,  .902} CodeLlama-7b & 0.95  & 0.95  & \textbf{-0.00}  & 6.28  & 5.15  & \textbf{-1.13}  & 9.74  & 7.87  & \textbf{-1.87}
    & 0.45  & 0.12  & \textbf{-0.33}  & 1.74 & 1.00  & \textbf{-0.74}  & 2.41  & 1.87  & \textbf{-0.54}  \\
    \rowcolor[rgb]{ .906,  .902,  .902} CodeLlama-13b & 3.02  & 1.97  & \textbf{-1.05}  & 16.39  & 10.29  & \textbf{-6.10}  & 23.22  & 14.98  & \textbf{-8.24}
    & 0.35  & 0.27  & \textbf{-0.08}  & 2.13 & 1.74  & \textbf{-0.39}  & 3.82  & 2.62  & \textbf{-1.20}  \\
    \rowcolor[rgb]{ .906,  .902,  .902} StarCoder & 0.80  & 0.50  & \textbf{-0.30}  & 5.09  & 3.64  & \textbf{-1.45}  & 7.89  & 6.37 & \textbf{-1.52}  
    & 0.10  & 0.05  & \textbf{-0.05} & 0.55  & 0.36  & \textbf{-0.19}  & 0.75  & 0.64  & \textbf{-0.11}  \\
    \rowcolor[rgb]{ .906,  .902,  .902} StarCoder2 & \textcolor{red}{\textbf{32.56}}  & \textcolor{red}{\textbf{32.43}}  & \textbf{-0.13}  & 50.98  & 50.61  & \textbf{-0.37}  & 55.81  & 55.43  & \textbf{-0.38}
    & 0.12  & 0.12  & \textbf{-0.00}  & 0.56 & 0.56  & \textbf{-0.00}  & 0.75  & 0.75  & \textbf{-0.00}  \\
    \rowcolor[rgb]{ .906,  .902,  .902} WizardCoder-15b-V1.0 & 8.49  & 5.17  & \textbf{-3.32}  & 25.32  & 16.18  & \textbf{-9.14}  & 29.59  & 19.48  & \textbf{-10.11}
    & 1.27  & 0.95  & \textbf{-0.32}  & 5.42 & 1.56  & \textbf{-3.86}  & 7.12  & 3.93  & \textbf{-3.19}  \\
    \midrule
    \multicolumn{1}{c||}{} & \multicolumn{9}{c||}{Java (High)}                                            & \multicolumn{9}{c}{C\# (Low)} \\
    \midrule
    Llama3-8b & 0.02  & 0.02  & \textbf{-0.00}  & 0.20  & 0.20  & \textbf{-0.00}  & 0.37  & 0.37  & \textbf{-0.00}
    & 0.22  & 0.07  & \textbf{-0.15}  & 1.78 & 0.60  & \textbf{-1.18}  & 3.33  & 1.12  & \textbf{-2.21}  \\
    Llama3.1-8b & 0.05  & 0.02  & \textbf{-0.03}  & 0.40  & 0.20  & \textbf{-0.20}  & 0.75  & 0.37 & \textbf{-0.38}
    & 1.09  & 0.95  & \textbf{-0.14}  & 14.59 & \textcolor{red}{\textbf{8.37}}  & \textbf{-6.22}  & 18.35  & \textcolor{red}{\textbf{11.24}}  & \textbf{-7.11}  \\
    Llama3.1-8b-Instruct & 0.52  & 0.30  & \textbf{-0.22}  & 2.72  & 1.46  & \textbf{-1.26}  & 3.75  & 1.87  & \textbf{-1.88} 
    & 1.11  & 0.92  & \textbf{-0.19}  & 8.00 & 4.81  & \textbf{-3.19}  & 13.33  & 7.12  & \textbf{-6.21}  \\
    Llama3.2-11b & 0.55  & 0.30  & \textbf{-0.25}  & 3.53  & 2.04  & \textbf{-1.49}  & 5.62  & 3.37  & \textbf{-2.25}
    & 1.56  & 1.06  & \textbf{-0.50}  & 10.19 & 3.71  & \textbf{-6.48}  & 16.67  & 4.12  & \underline{\textbf{-12.55}}  \\
    Phi3-medium-4k-instruct & 2.82  & 1.55  & \textbf{-1.27}  & 9.83  & 7.03  & \textbf{-2.80}  & 11.99  & \textcolor{red}{\textbf{9.36}}  & \textbf{-2.63} 
    & 1.90  & 1.32  & \textbf{-0.58}  & 8.95 & 6.11  & \textbf{-2.84}  & 11.99  & 8.61  & \textbf{-3.38}  \\
    Qwen2.5-7b-Instruct & 3.67  & 0.77  & \textbf{-2.90}  & 13.48  & 4.51  & \textbf{-8.94}  & 16.85  & 6.74  & \textbf{-10.11}
    & 1.87  & 0.95  & \underline{\textbf{-0.92}}  & 10.08 & 5.88  & \textbf{-4.20}  & 14.23  & 9.36  & \textbf{-4.87}  \\
    Qwen2.5-14b-Instruct & 5.42  & 1.32  & \underline{\textbf{-4.10}}  & \textcolor{red}{\textbf{16.82}}  & 6.74  & \underline{\textbf{-10.08}}  & \textcolor{red}{\textbf{19.85}}  & \textcolor{red}{\textbf{9.36}}  & \underline{\textbf{-10.49}} 
    & \textcolor{red}{\textbf{2.13}}  & \textcolor{red}{\textbf{1.84}}  & \textbf{-0.29}  & \textcolor{red}{\textbf{18.18}} & 7.91  & \underline{\textbf{-10.27}}  & \textcolor{red}{\textbf{20.08}}  & 9.56  & \textbf{-10.52}  \\
    Vicuna-13b-v1.5 & 0.15  & 0.02  & \textbf{-0.13}  & 0.85  & 0.20  & \textbf{-0.65}  & 1.12  & 0.37  & \textbf{-0.75}
    & 0.12  & 0.05  & \textbf{-0.07}  & 1.00  & 0.40  & \textbf{-0.60}  & 1.87  & 0.75 & \textbf{-1.12}  \\
    \rowcolor[rgb]{ .906,  .902,  .902} CodeLlama-7b & 0.22  & 0.07  & \textbf{-0.15}  & 1.60  & 0.60  & \textbf{-1.00}  & 2.62  & 1.12  & \textbf{-1.50}
    & 0.25  & 0.12  & \textbf{-0.13}  & 1.80 & 1.00  & \textbf{-0.80}  & 3.00  & 1.87  & \textbf{-1.13}  \\
    \rowcolor[rgb]{ .906,  .902,  .902} CodeLlama-13b & 2.62  & 0.13  & \textbf{-2.46}  & 7.03  & 2.07  & \textbf{-4.96}  & 8.24  & 3.12  & \textbf{-5.12}
    & 0.30  & 0.27  & \textbf{-0.03}  & 2.20 & 1.74  & \textbf{-0.46}  & 3.75  & 2.62  & \textbf{-1.13}  \\
    \rowcolor[rgb]{ .906,  .902,  .902} StarCoder & 0.02  & 0.00  & \textbf{-0.02}  & 0.27  & 0.10  & \textbf{-0.17}  & 0.45  & 0.18  & \textbf{-0.27}
    & 0.40  & 0.32  & \textbf{-0.08}  & 4.52 & 2.09  & \textbf{-2.43}  & 6.64  & 4.74  & \textbf{-1.90}  \\
    \rowcolor[rgb]{ .906,  .902,  .902} StarCoder2 & \textcolor{red}{\textbf{7.57}}  & \textcolor{red}{\textbf{5.72}}  & \textbf{-1.85}  & 12.66  & \textcolor{red}{\textbf{9.32}}  & \textbf{-3.34}  & 13.48  & 10.11  & \textbf{-3.37}
    & 1.20  & 0.82  & \textbf{-0.38}  & 5.99 & 3.98  & \textbf{-2.01}  & 7.87  & 5.24  & \textbf{-2.63}  \\
    \rowcolor[rgb]{ .906,  .902,  .902} WizardCoder-15b-V1.0 & 1.47  & 1.07  & \textbf{-0.40}  & 4.92  & 3.67  & \textbf{-1.25}  & 5.62  & 4.12  & \textbf{-1.50}
    & 1.50  & 0.97  & \textbf{-0.53}  & 6.16 & 3.82  & \textbf{-2.34}  & 8.61  & 5.24  & \textbf{-3.37}  \\
    \midrule
    \multicolumn{1}{c||}{} & \multicolumn{9}{c||}{PHP (Low)}                                            & \multicolumn{9}{c}{JavaScript (Medium)} \\
    \midrule
    Llama3-8b & 1.17  & 0.82  & \textbf{-0.35}  & 8.74  & 6.29  & \textbf{-2.45}  & 15.36  & 10.86  & \textbf{-4.50}
    & 0.37  & 0.25  & \textbf{-0.12}  & 2.90 & 1.90  & \textbf{-1.00}  & 5.24  & 3.37  & \textbf{-1.87}  \\
    Llama3.1-8b & 2.42  & 2.00  & \textbf{-0.42}  & 17.16  & \textcolor{red}{\textbf{14.67}}  & \textbf{-2.49}  & 29.21  & \textcolor{red}{\textbf{24.47}}  & \textbf{-4.74}
    & 0.20  & 0.12  & \textbf{-0.08}  & 1.50 & 1.00  & \textbf{-0.50}  & 2.62  & 1.87  & \textbf{-0.75}  \\
    Llama3.1-8b-Instruct & 6.77  & 0.72  & \textbf{-6.02}  & 28.96  & 4.02  & \textbf{-24.94}  & 37.83  & 5.99  & \textbf{-31.84} 
    & 4.17  & 3.05  & \textbf{-1.12}  & 19.04 & 12.82  & \textbf{-6.22}  & 25.84  & 17.23  & \textbf{-8.61}  \\
    Llama3.2-11b & 7.54  & 1.05  & \textbf{-6.49}  & 30.62  & 2.05  & \textbf{-28.57}  & 39.33  & 2.25  & \textbf{-37.08}
    & 3.22  & 2.27  & \textbf{-0.95}  & 15.93 & 10.49  & \textbf{-5.44}  & 21.72  & 13.48  & \textbf{-8.24}  \\
    Phi3-medium-4k-instruct & 14.11  & 1.77  & \textbf{-12.34}  & 42.49  & 7.48  & \textbf{-35.01}  & 50.19  & 10.11  & \textbf{-40.08} 
    & 8.61  & 5.37  & \textbf{-3.24}  & 32.24 & 22.33  & \textbf{-9.91}  & 41.20  & 29.21  & \textbf{-11.99}  \\
    Qwen2.5-7b-Instruct & 14.08  & 1.00  & \textbf{-13.08}  & 47.56  & 4.95  & \textbf{-42.61}  & 57.68 & 6.37  & \textbf{-51.31}
    & 13.43  & 6.69  & \textbf{-6.74}  & 48.02 & 28.14  & \underline{\textbf{-19.88}}  & 56.93  & 35.96  & \underline{\textbf{-20.97}}  \\
    Qwen2.5-14b-Instruct & \textcolor{red}{\textbf{22.22}}  & \textcolor{red}{\textbf{2.12}}  & \underline{\textbf{-20.10}}  & \textcolor{red}{\textbf{63.56}}  & 7.25  & \underline{\textbf{-56.31}}  & \textcolor{red}{\textbf{73.41}}  & 8.24  & \underline{\textbf{-65.17}} 
    & \textcolor{red}{\textbf{16.13}}  & \textcolor{red}{\textbf{8.54}}  & \underline{\textbf{-7.59}}  & \textcolor{red}{\textbf{52.52}} & \textcolor{red}{\textbf{35.51}}  & \textbf{-17.01}  & \textcolor{red}{\textbf{61.05}}  & \textcolor{red}{\textbf{44.19}}  & \textbf{-16.86}  \\
    Vicuna-13b-v1.5 & 0.22  & 0.05  & \textbf{-0.17}  & 1.44  & 0.40  & \textbf{-1.04}  & 2.25  & 0.75  & \textbf{-1.50}
    & 0.27  & 0.22  & \textbf{-0.05}  & 1.69 & 1.29  & \textbf{-1.40}  & 2.62  & 1.87  & \textbf{-0.75}  \\
    \rowcolor[rgb]{ .906,  .902,  .902} CodeLlama-7b & 1.40  & 0.85  & \textbf{-0.55}  & 9.98  & 6.39  & \textbf{-3.59}  & 16.85  & 11.24  & \textbf{-5.61}
    & 0.50  & 0.15  & \textbf{-0.35}  & 3.70 & 1.10  & \textbf{-2.60}  & 6.37  & 1.87  & \textbf{-4.50}  \\
    \rowcolor[rgb]{ .906,  .902,  .902} CodeLlama-13b & 1.32  & 0.55  & \textbf{-0.77}  & 9.38  & 3.94  & \textbf{-5.44}  & 13.18  & 6.74  & \textbf{-6.44}
    & 0.82  & 0.35  & \textbf{-0.47}  & 5.99 & 2.60  & \textbf{-3.39}  & 10.11  & 4.49  & \textbf{-5.62}  \\
    \rowcolor[rgb]{ .906,  .902,  .902} StarCoder & 0.12  & 0.05  & \textbf{-0.07}  & 0.90  & 0.40  & \textbf{-0.50}  & 1.50  & 0.75  & \textbf{-0.75}
    & 0.07  & 0.07  & \textbf{-0.00}  & 0.60 & 0.60  & \textbf{-0.00}  & 1.12  & 1.12  & \textbf{-0.00}  \\
    \rowcolor[rgb]{ .906,  .902,  .902} StarCoder2 & 5.57  & 0.32  & \textbf{-5.25}  & 28.42  & 2.34  & \textbf{-26.08}  & 39.70  & 4.12  & \textbf{-35.58}
    & 8.69  & 5.02  & \textbf{-3.67}  & 31.78 & 17.20  & \textbf{-14.58}  & 38.58  & 20.22  & \textbf{-18.36}  \\
    \rowcolor[rgb]{ .906,  .902,  .902} WizardCoder-15b-V1.0 & 6.27  & 1.50  & \textbf{-4.77}  & 21.99  & 4.74  & \textbf{-17.52}  & 26.97  & 5.62  & \textbf{-21.35}
    & 5.49  & 3.12  & \textbf{-2.37}  & 19.07 & 10.95  & \textbf{-8.12}  & 23.97  & 14.23  & \textbf{-9.74}  \\
    \bottomrule
    \end{tabular}%
    }
  \label{tab:result_02}%
\end{table*}%

\begin{figure*}[!t]
    \centering
    \begin{subfigure}{0.47\textwidth}
        \centering
        \includegraphics[width=\linewidth]{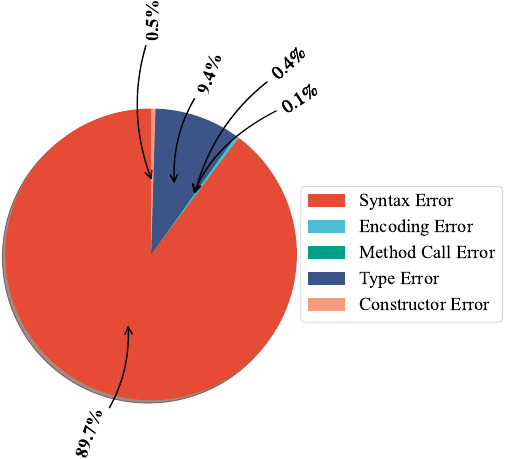}
        \caption{Java}
    \end{subfigure}
    \hspace{0.005cm}
    \begin{subfigure}{0.47\textwidth}
        \centering
        \includegraphics[width=\linewidth]{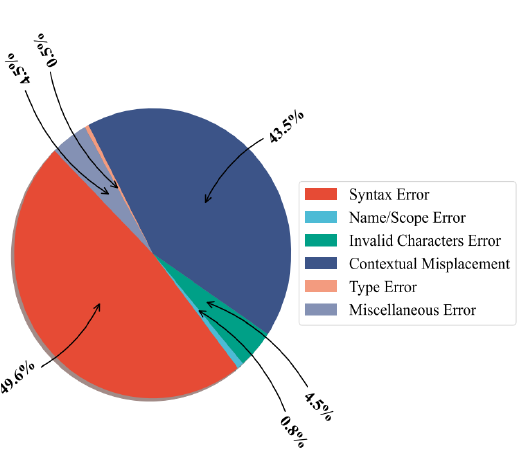}
        \caption{C\#}
    \end{subfigure}
    \begin{subfigure}{0.47\textwidth}
        \centering
        \includegraphics[width=\linewidth]{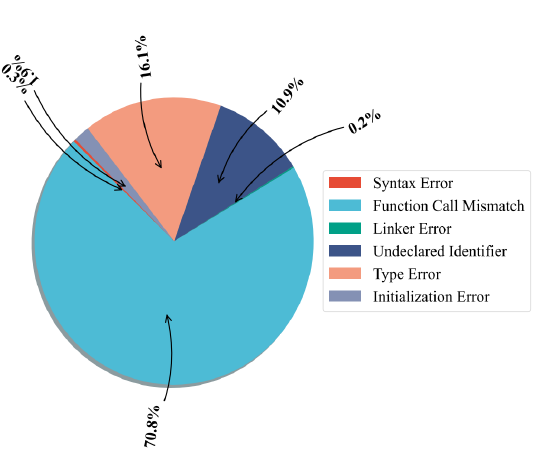}
        \caption{PHP}
    \end{subfigure}
    \begin{subfigure}{0.47\textwidth}
        \centering
        \includegraphics[width=\linewidth]{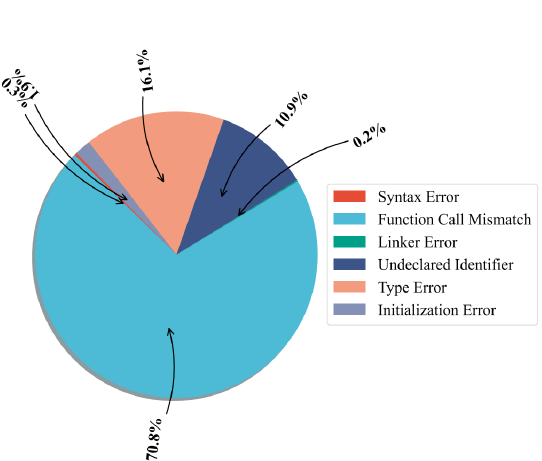}
        \caption{JavaScript}
    \end{subfigure}
    \caption{Error distribution of Java, C\#, PHP and JavaScript programming languages.}
    \label{fig:error_distribution_other}%
\end{figure*}

\section{Supplementary results}
As clearly shown in Table~\ref{tab:result_02}, most LLMs exhibit generally poor performance on the MultiOOP benchmark, with substantial variation across programming languages. Notably, their performance on high-frequency languages (such as Python and C++) is not necessarily superior to that on low-frequency languages (such as PHP and C\#). Furthermore, LLMs continue to struggle with understanding and implementing core OOP concepts and features, such as classes and inheritance. This is particularly evident in their inconsistent and inaccurate execution of user-specified requirements, indicating that they have yet to fully meet expectations in this domain.

\section{Supplementary Bad Case Analysis}
This section first analyzes the reasons why LLMs make errors when performing OOP tasks in Python and C++, as illustrated in Listing~\ref{lst:error_python} and~\ref{lst:error_C}. We then further supplement this analysis with the distribution and examples of errors made by the LLM in other languages, including Java, C\#, PHP, and JavaScript, also detailed in Figure~\ref{fig:error_distribution_other}.

\noindent
\textbf{Possible reasons for LLMs failing in OOP tasks in Python and C++ language.} As shown clearly in Listing~\ref{lst:error_python}, the code generated by the StarCoder2 fails to follow the naming conventions specified in the input (i.e., using ``Solution'' instead of ``LCMP'', and ``longestCommonPrefix'' instead of ``longest\_common\_prefix''. This phenomenon indicates that StarCoder2 has certain limitations in understanding and applying specific naming rules. This may be attributed to an overreliance on common naming patterns in its training data, such as the frequent use of the ``Solution'' class name in LeetCode~\footnote{\url{https://leetcode.com/.}} problems.
As shown clearly in Listing~\ref{lst:error_C}, the code generated by the Qwen2.5-7b-Instruct provides only a skeletal framework and lacks concrete functional implementation. This indicates that the Qwen2.5-7b-Instruct may have been trained on a large corpus of templated or boilerplate code, particularly from educational sources or programming platforms e.g., LeetCode, which may lead it to emphasize structural elements such as class names and function signatures, while omitting core algorithmic logic. Despite the user's explicit request to solve the ``trapping rainwater'' problem, the Qwen2.5-7b-Instruct merely extracted superficial components such as the class name (``HTRW'') and function name (``harvest\_rainwater''), failing to understand and implement the required algorithm. This phenomenon indicates that Qwen2.5-7b-Instruct tends to overemphasize formal structure when confronted with complex problem descriptions, often at the expense of substantive functional implementation.

\noindent
\textbf{Error distribution of LLMs in OOP tasks across other programming languages.}
As illustrated in Figure~\ref{fig:error_distribution_other}, it is evident that LLMs demonstrate distinct categories of prevalent errors across different programming languages within OOP tasks. In the case of Java language, the primary error types include ``\textit{Syntax Error}'', ``\textit{Encoding Error}'', ``\textit{Constructor Error}'', ``\textit{Type Error}'', and ``\textit{Method Call Error}'', with ``\textit{Syntax Error}'' comprising as much as 89.7\% of the total. For C\# language, frequently observed errors encompass ``\textit{Syntax Error}'', ``\textit{Invalid Characters Error}'', ``\textit{Contextual Misplacement}'', ``\textit{Type Error}'', ``\textit{Name/Scope Error}'', as well as various ``\textit{Miscellaneous Error}''. Regarding PHP language, common error types consist of ``\textit{Syntax Error}'', ``\textit{Function Call Mismatch}'', ``\textit{Undeclared Identifier}'', ``\textit{Type Error}'', ``\textit{Initialization Error}'', and ``\textit{Linker Error}''. Lastly, in JavaScript language, typical errors include ``\textit{Syntax Error}'', ``\textit{Runtime Error}'', ``\textit{Reference Error}'', ``\textit{Type Error}'' and ``\textit{Toolchain Error}''. For common error examples made by LLMs in Java and C\#, as shown in Listing~\ref{lst:error_Java} and~\ref{lst:error_cs}.

The above errors indicate that LLMs have an insufficient grasp of the syntax rules of different programming languages when handling MultiOOP tasks, and they also reflect the LLMs’ limitations in understanding and applying the unique features of each language. Furthermore, these mistakes reveal shortcomings in the LLMs’ ability to follow user instructions. For example, they may correctly recognize class names and function signatures, yet fail to implement the required logic or adhere to naming conventions.



\begin{lstlisting}[language=Python, escapeinside={(*@}{@*)}, caption={An error example produced by StarCoder2, where it fails to generate the required class name and public function name specified by the user, resulting in an AssertionError.}, label={lst:error_python}]
(*@\textbf{Input:}@*)
First, write a (*@\textbf{LCMP}@*) class using the Python language. 
Then, within the  (*@\textbf{LCMP}@*) class, create a public 
function called (*@\textbf{longest\_common\_prefix}@*) to find the 
longest common prefix among an array of strings. If 
no common prefix exists, return an empty string.

(*@\textbf{Python code:}@*)
class Solution:
    def longestCommonPrefix(self, strs: List[str]) -> str:
        if not strs:
            return ''
        prefix = strs[0]
        for i in range(1, len(strs)):
            while strs[i].find(prefix) != 0:
                prefix = prefix[:-1]
        return prefix
\end{lstlisting}

\begin{lstlisting}[language=C++, escapeinside={(*@}{@*)}, caption={An example of an error produced by Qwen2.5-7b-Instruct, where only the framework of the user's requirements is defined without implementing the specific functionality, resulting in a syntax or compile-time error.}, label={lst:error_C}]
(*@\textbf{Input:}@*)
First, write an (*@\textbf{HTRW}@*) class using the C++ language, 
then write a public function named
(*@\textbf{harvest\_rainwater}@*) within the (*@\textbf{HTRW}@*) class to solve the 
following problem.

Problem: Given (*@\textbf{n}@*) non-negative integers representing 
the height of each pillar of width 1 in the diagram,
calculate how much rainwater can be collected after 
it rains with the pillars arranged in this way.

(*@\textbf{C++ code:}@*)
class HTRW {
public:
    // Function to calculate the amount of rainwater that can be harvested
    int harvest_rainwater(vector<int>& height) {
        // Your implementation goes here
    }
};
\end{lstlisting}

\begin{lstlisting}[language=Java, escapeinside={(*@}{@*)}, caption={An example of an error generated by StarCoder2. The error involves a ``\textit{Type Error}'' in StarCoder2.}, label={lst:error_Java}]
(*@\textbf{Input:}@*)
First, write a (*@\textbf{CRTP}@*) class using the Java language. 
Then, within the (*@\textbf{CRTP}@*) class, implement a public 
function called (*@\textbf{climb\_rooftop}@*) to solve the following 
problem:
Suppose you are climbing a staircase and it takes (*@\textbf{n}@*) 
steps to reach the top. At each step, you can either 
climb 1 or 2 steps. How many distinct ways are there 
to climb to the top?

(*@\textbf{Java code:}@*)
public class ClimbStairs<T extends ClimbStairs<T>> {
    public int climb_rooftop(int n) {
        if (n == 1) {
            return 1;
        } else if (n == 2) {
            return 2;
        } else {
            return ((T) this).climb_rooftop(n - 1) + ((T) this).climb_rooftop(n - 2);
        }
    }
}
\end{lstlisting}

\begin{lstlisting}[language={C++}, escapeinside={(*@}{@*)}, caption={An example of an error generated by Phi3-medium-4k-instruct. The error involves ``\textit{Syntax Error}'', ``\textit{Tye Error}'' in Phi3-medium-4k-instruct.}, label={lst:error_cs}]
(*@\textbf{Input:}@*)
First, write a (*@\textbf{CRTP}@*) class using the Java language. 
Then, within the (*@\textbf{CRTP}@*) class, implement a public 
function called (*@\textbf{climb\_rooftop}@*) to solve the following 
problem:
Suppose you are climbing a staircase and it takes (*@\textbf{n}@*) 
steps to reach the top. At each step, you can either 
climb 1 or 2 steps. How many distinct ways are there 
to climb to the top?

(*@\textbf{C\# code:}@*)
public class Staircase<T> where T : Staircase<T>, new()
{
    public T ClimbRooftop(int n)
    {
        T instance = new T();
        int[] ways = new int[n + 1];
        ways[0] = 1;
        ways[1] = 1;

        for (int i = 2; i <= n; i++)
        {
            ways[i] = ways[i - 1] + ways[i - 2];
        }

        instance.DifferentWaysToClimb = ways[n];
        return instance;
    }
}

public class StaircaseSolution : Staircase<StaircaseSolution>
{
    public int DifferentWaysToClimb { get; private set; }

    public StaircaseSolution(int n)
    {
        DifferentWaysToClimb = ClimbRooftop(n).DifferentWaysToClimb;
    }
}
\end{lstlisting}

\section{Design considerations for the number of test cases in the MultiOOP benchmark}
\begin{figure}[!t]
    \centering
    \includegraphics[width=0.45\textwidth]{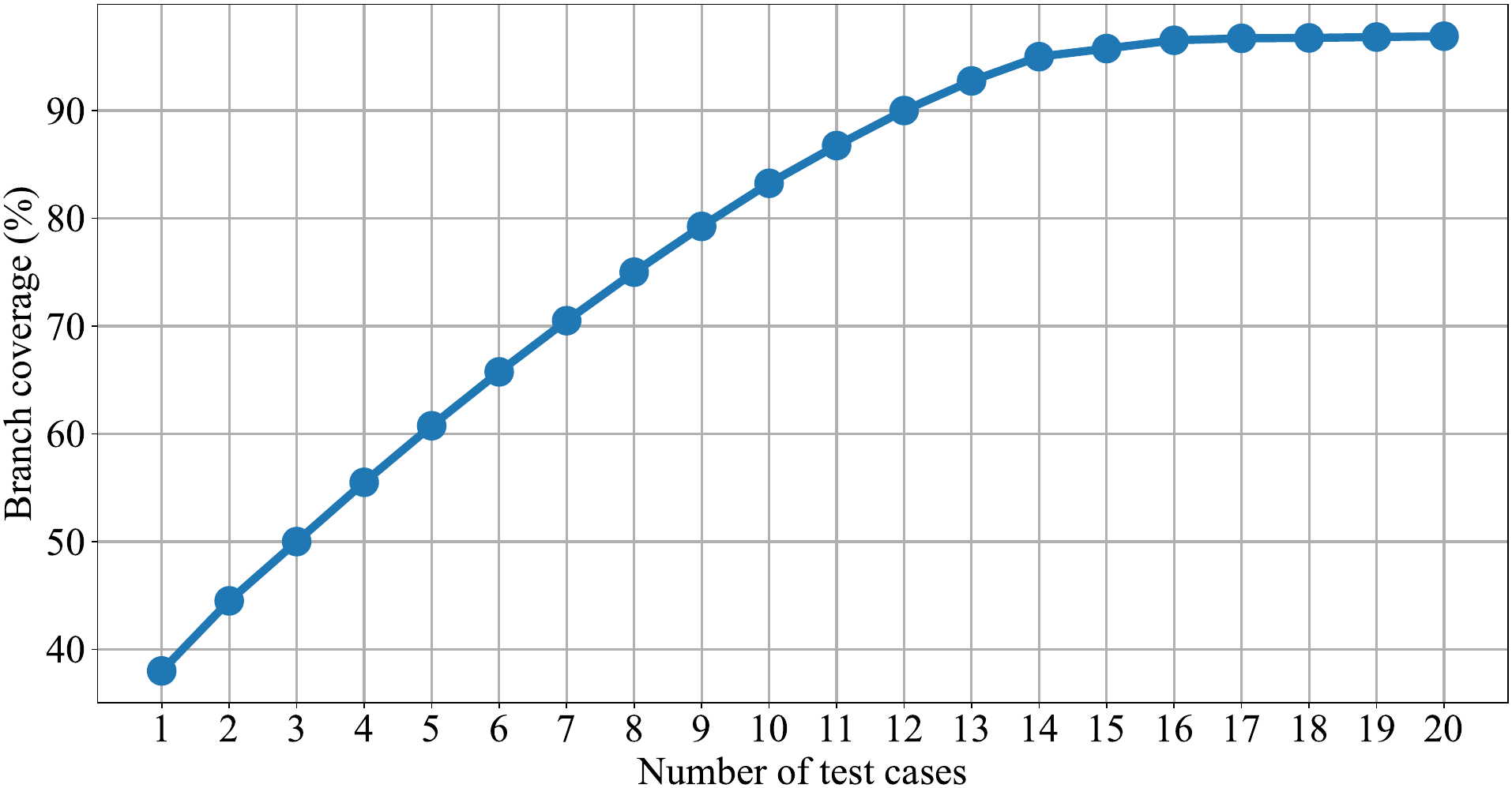}
    \caption{The changes in branch coverage with the increasing number of test cases.}
    \label{fig:branch_coverage}
\end{figure}

According to the results shown in Figure $6$, when the number of test cases is fewer than $15$, the performance of the LLM decreases; however, once the number of test cases reaches $15$, the LLM’s performance gradually stabilizes. Subsequently, we incrementally increased the number of test cases and applied them to evaluate the ground-truth Python code. As observed in Figure~\ref{fig:branch_coverage}, when the number of test cases reaches $18$, the branch coverage levels off. Therefore, in the MultiOOP benchmark, we maintain an average of $20$ test cases. Note: Here, we use the coverage tool~\footnote{~\url{https://pypi.org/project/coverage/.}} to evaluate branch coverage.

\end{document}